\documentclass[10pt,prl,aps,superscriptaddress,twocolumn,notitlepage,noeprint]{revtex4-1}
\usepackage{amssymb,amsmath}
\usepackage{graphicx}
\usepackage{color}
\usepackage{natbib}
\usepackage{hyperref}
\usepackage[utf8]{inputenc}

\usepackage[normalem]{ulem}
\usepackage{siunitx}
\usepackage[russian,english]{babel}

\begin{document}

\newcommand{\e}{{\rm e}}
\newcommand{\norm}[1]{\left\lVert#1\right\rVert}
\newcommand{\rmi}{{\rm i}}
\renewcommand{\Im}{\mathop\mathrm{Im}\nolimits}
\renewcommand{\Re}{\mathop\mathrm{Re}\nolimits}
\newcommand{\red}[1]{{\color{red}#1}}
\newcommand{\purple}[1]{{\color{black}#1}}
\newcommand{\blue}[1]{{\color{blue}#1}}
\newcommand{\rot}[1]{\text{rot}\, #1}
\newcommand{\df}[2]{\frac{\partial #1}{\partial #2}}
\newcommand{\bas}[1]{\hat{{\bf #1}}}
\renewcommand{\cite}[1]{[\onlinecite{#1}]}

\newcommand{\ket}[1]{\left|#1\right>}      
\newcommand{\bra}[1]{\left<#1\right|}
\newcommand{\eps}{\varepsilon}      
\newcommand{\om}{\omega}      
\newcommand{\kap}{\varkappa}      
\newcommand{\VB}[1]{\mathbf{#1}} 
\newcommand{\n}{\nabla}
\newcommand{\sgn}{sgn} 

\renewcommand{\thefootnote}{\roman{footnote}}

\newcommand{\affilITMO}{School of Physics and Engineering, ITMO University,  St.-Petersburg 197101, Russia}

\title{Broadband chiroptical sensing with two-mode subwavelength grating  waveguides}

\author{Albert~A.~Seredin}
\affiliation{\affilITMO}

\author{Roman~S.~Savelev}
\affiliation{\affilITMO}

\begin{abstract}
Engineering local optical chirality in planar photonic structures offers a route to selective and enhanced light–matter interactions with chiral media. Here, we introduce periodic dielectric waveguides that support chiral evanescent fields through the interaction of two nearly degenerate waveguide modes. We develop a perturbation theory for such a waveguide covered by a chiral layer, relating chirality-induced modal coupling and propagation-constant corrections to measurable intermodal power conversion and differential transmission. Whereas modal detuning confines the chiral response of a symmetric waveguide to a narrow spectral range, mode hybridization in an asymmetric waveguide creates individual eigenmodes that carry nonzero local chirality over a broad spectral range. We show that modal dichroism in realistic lossy waveguides can produce a significant background differential transmission signal. Such a background, however, can be suppressed by comparing transmission through two asymmetric waveguides with opposite structural asymmetries, allowing the chiral response to be reliably extracted. Together, these findings provide a basis for developing integrated photonic components with locally chiral fields, including waveguides, metasurfaces, and resonators.
\end{abstract}

\maketitle

\section{Introduction}

Chirality is a fundamental property of many molecules and molecular assemblies in chemistry and biology, with opposite enantiomers often exhibiting markedly different chemical and biological functions despite having identical compositions~\cite{LiuNature2023}. Reliable detection and discrimination of molecular handedness and enantiomeric composition are therefore important in chemistry, pharmaceutical science, and biomedical diagnostics~\cite{Hendry2010,LiuNature2023,SolomonACSPhotonics2019}.

Optical activity and circular dichroism (CD) provide label-free approaches for probing chirality through the differential interaction of a sample with electromagnetic fields of opposite handedness~\cite{TangScience2011}. In the mid-infrared, vibrational circular dichroism (VCD) is particularly attractive because it combines enantiosensitivity with the molecular specificity of vibrational transitions, providing direct access to molecular structure and conformation~\cite{FreedmanChirality2003,KiederlingMolecules2018,PescitelliJPCB2024}. Its major limitation, however, is the intrinsically weak chiroptical response, several orders weaker than the corresponding infrared absorption signals, making measurements of dilute or nanoscale samples particularly challenging~\cite{KiederlingMolecules2018,HermannAnalChem2022,PescitelliJPCB2024}.

At the microscopic level, the local handedness of the electromagnetic field can be characterized by the optical chirality density, which for monochromatic fields (assuming $e^{-i\omega t}$ time dependence) can be written as~\cite{Tang2010}:
\begin{equation}
C(\mathbf{r}) = \dfrac{k_0}{2c}\Im[\VB{E}(\VB{r})\cdot\VB{H}^*(\VB{r})],
\label{eq:C_def}
\end{equation}

For a circularly polarized plane wave, the optical chirality is directly related to the electromagnetic energy density and provides a natural reference value. Nanophotonic structures offer additional control over the magnitude, sign, and spatial distribution of optical chirality, enabling chiral fields to be concentrated in regions accessible to molecules. These capabilities have motivated a broad range of approaches for enhancing weak molecular chiroptical responses.

For molecular CD enhancement, both field helicity and the spatial distribution of optical chirality across the sample must be considered. In particular, helicity-preserving designs help avoid cancellation between contributions of opposite chirality~\cite{SolomonACSPhotonics2019,GrafACSPhotonics2019,LasaACSPhotonics2020}. Early approaches relied extensively on plasmonic nanostructures, where strong subwavelength field confinement was used to generate enhanced chiral hot spots and amplify molecular chiroptical signals~\cite{Hendry2010,SchafferlingPRX2012,WarningACS2021}. All-dielectric nanostructures and metasurfaces subsequently emerged as a low-loss alternative, with electric and magnetic resonances that can be engineered to produce enhanced optical chirality of predominantly uniform sign in accessible sensing regions~\cite{MohammadiACSPhotonics2019,RazimanACSPhotonics2019}. Experimental studies have demonstrated enhanced CD signals for molecular coatings deposited on arrays of dielectric nanoresonators~\cite{GarciaGuiradoNanoLett2020,SolomonJACS2020}. High-$Q$ resonances can further improve chiral sensitivity by increasing the photon lifetime, while narrowing the operational spectral range. Related realizations include high-$Q$ and bound-state-in-the-continuum metasurfaces, helicity-preserving optical cavities, and whispering-gallery-mode resonators~\cite{HuACSPhotonics2020,ChenNanoLett2020,FeisPRL2020,TianOE2024,ShakirovaACS2025,NouriLPR2026}.
These concepts are directly applicable to vibrational spectroscopy: germanium dielectric metasurfaces operating in the mid-infrared spectral range have been proposed to generate strongly enhanced optical chirality within thin molecular layers, providing a route toward enhanced VCD spectroscopy~\cite{YeACS2023}.

Optical dielectric waveguides provide a complementary route to chiral sensing by extending the light–matter interaction over long propagation distances, as recently demonstrated in optical-rotation measurements using liquid-filled hollow-core fibers~\cite{HelfrichJPP2026}. In integrated waveguides, interaction lengths spanning many thousands of wavelengths can be realized on a planar substrate, while the evanescent field remains accessible to liquids and molecular films, making these structures naturally compatible with integrated sensing and microfluidic architectures. These advantages have already enabled highly sensitive refractive-index and absorption measurements~\cite{LuanSensors2018,PerezOLT2019,WangAMT2020,XiaoAOM2023}. The electromagnetic fields of dielectric waveguides can also possess substantial local chirality due to the vectorial nature of the field components of guided modes~\cite{PicardiOPT2018}. Recently, it has been shown that strongly chiral fields can be generated through coherent superpositions of quasi-degenerate guided modes~\cite{PellegriniPRB2017,VazquezLozano2019,RomeuNanophotonics2024}, opening the possibility of implementing chiral sensing and spectroscopy on planar photonic integrated platforms.

\begin{figure}[t]
    \centering
    \includegraphics[width=\columnwidth]
{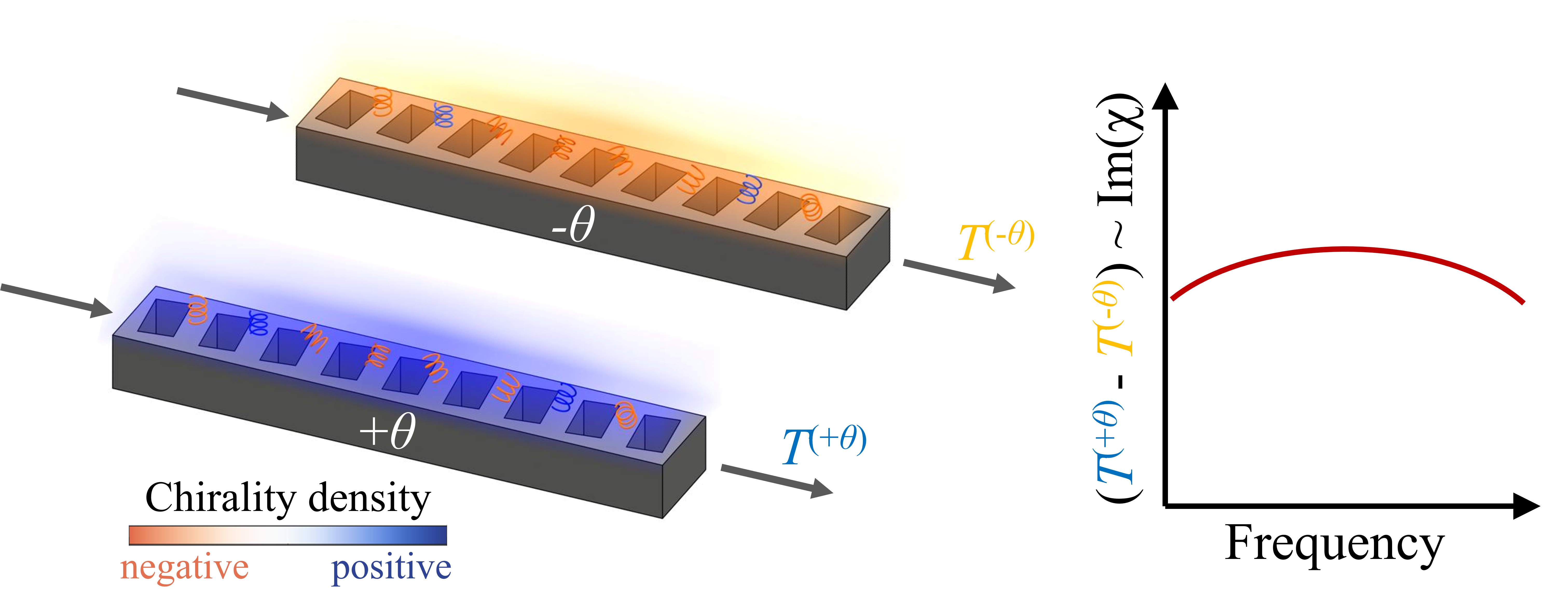}
    \caption{Schematic of broadband chiral sensing using a pair of periodic two-mode waveguides with opposite structural asymmetries, $+\theta$ and $-\theta$.
    The corresponding chiral fields interact with the same chiral medium and produce transmissions $T^{(+\theta)}$ and $T^{(-\theta)}$, which difference is linearly proportional to the chirality parameter over a broad spectral range.}
    \label{fig:concept}
\end{figure}

Despite this progress, two important practical limitations remain. First, strongly chiral guided fields based on coherent superpositions require a well-defined relative amplitude and phase between the participating modes. In realistic waveguides, chiral superposition can be maintained over an appreciable interaction length only near a modal degeneracy or within a narrow spectral interval, restricting the bandwidth available for chiral spectroscopy~\cite{VazquezLozano2019,RomeuJOSAb2026}. Second, the generation of a locally chiral field does not by itself determine the measurable response of a waveguide covered by a thin chiral layer. A quantitative sensing model must relate the spatial overlap of the layer with the modal optical chirality to the chirality-induced corrections to the modal propagation constants and, ultimately, to differential attenuation or transmission~\cite{BothACSNano2022,EnriquezJOSAb2023,AlmousaPRB2024}. Establishing this connection is particularly important for VCD spectroscopy, which requires both broadband operation and a measurable response from thin molecular films.

In this work, we address these limitations by engineering the coupling of two guided modes in a periodic dielectric waveguide. We first use a cylindrical-waveguide model to show how a coherent superposition of two modes generates optical chirality in the evanescent field, despite the achirality of the individual modes and the underlying structure. Guided by this model, we develop realistic suspended dielectric waveguides for the mid-infrared spectral range. A controlled in-plane perturbation couples the relevant modes, producing eigenmodes with the required relative modal amplitude and phase over a broad spectral range, without requiring a geometrically chiral or twisted waveguide, as schematically illustrated in Fig.~\ref{fig:concept}. The normalized optical chirality in the upper half-space remains of order unity even after averaging over a molecular layer of finite thickness. Finally, we derive a perturbative framework that relates the optical chirality sampled by the layer to chirality-induced changes in the complex modal propagation constants and the corresponding differential transmission. These results establish a direct connection between geometry-controlled modal coupling, local optical chirality, and an experimentally accessible chiroptical signal, providing a promising platform for broadband on-chip chiral sensing and vibrational circular dichroism spectroscopy.

\section{Analytical toy model}

To elucidate the basic physical mechanism of chirality generation through the interference of guided modes and to obtain simple analytical results, we consider a simplified toy model: a dielectric cylindrical waveguide of radius $R$ with its axis along the $z$ direction. The cladding is assumed to be air, with $\varepsilon=1$ and $\mu=1$. We focus on the TE$_{01}$ and HE$_{11}^y$ modes, which possess the same parity with respect to reflection in the plane $y=0$.

For simplicity, we approximate the radial dependence of the fields in the cladding region ($r>R$) as
\begin{equation}
F_j(r)=\frac{\exp[-\gamma_j(r-R)]}{\sqrt{r/R}},
\qquad
\gamma_j=\sqrt{\beta_j^2-k_0^2},
\end{equation}
where $j=1,2$ denotes the two modes, $k_0=\omega/c$, and $c$ is the speed of light. We adopt the time and longitudinal dependence $\exp[i(\beta_jz-\omega t)]$ and write $A_j(r)=a_jF_j(r)$, where $a_j$ is a modal normalization constant. Within this approximation, the field components of the TE$_{01}$ mode can be written as~\cite{Okamoto}
\begin{equation}
E_\phi = A_1,\; Z_0H_r = -\widetilde{\beta}_1 A_1,\; Z_0H_z = i\widetilde{\gamma}_1A_1,
\label{eq:TE_clad}
\end{equation}
whereas those of the HE$_{11}^y$ mode are
\begin{equation}
\begin{aligned}
E_r &= A_2\sin\phi, \; Z_0H_r = -\widetilde{\beta}_2A_2\cos\phi, \\ E_\phi &= A_2\cos\phi, \; Z_0H_\phi = \frac{1}{\widetilde{\beta}_2}A_2\sin\phi, \\
E_z &= -i\frac{\widetilde{\gamma}_2}{\widetilde{\beta}_2}A_2\sin\phi, \; Z_0H_z = i\widetilde{\gamma}_2A_2\cos\phi.
\end{aligned}
\label{eq:HE_clad}
\end{equation}
Here, $\widetilde{\beta}_j=\beta_j/k_0$, $\widetilde{\gamma}_j=\gamma_j/k_0$, and $Z_0$ is the free-space impedance. Details of the fields approximation are provided in Supporting Information, Sec.~I.

\begin{figure}[t]
    \center{\includegraphics[width=1\columnwidth]
    {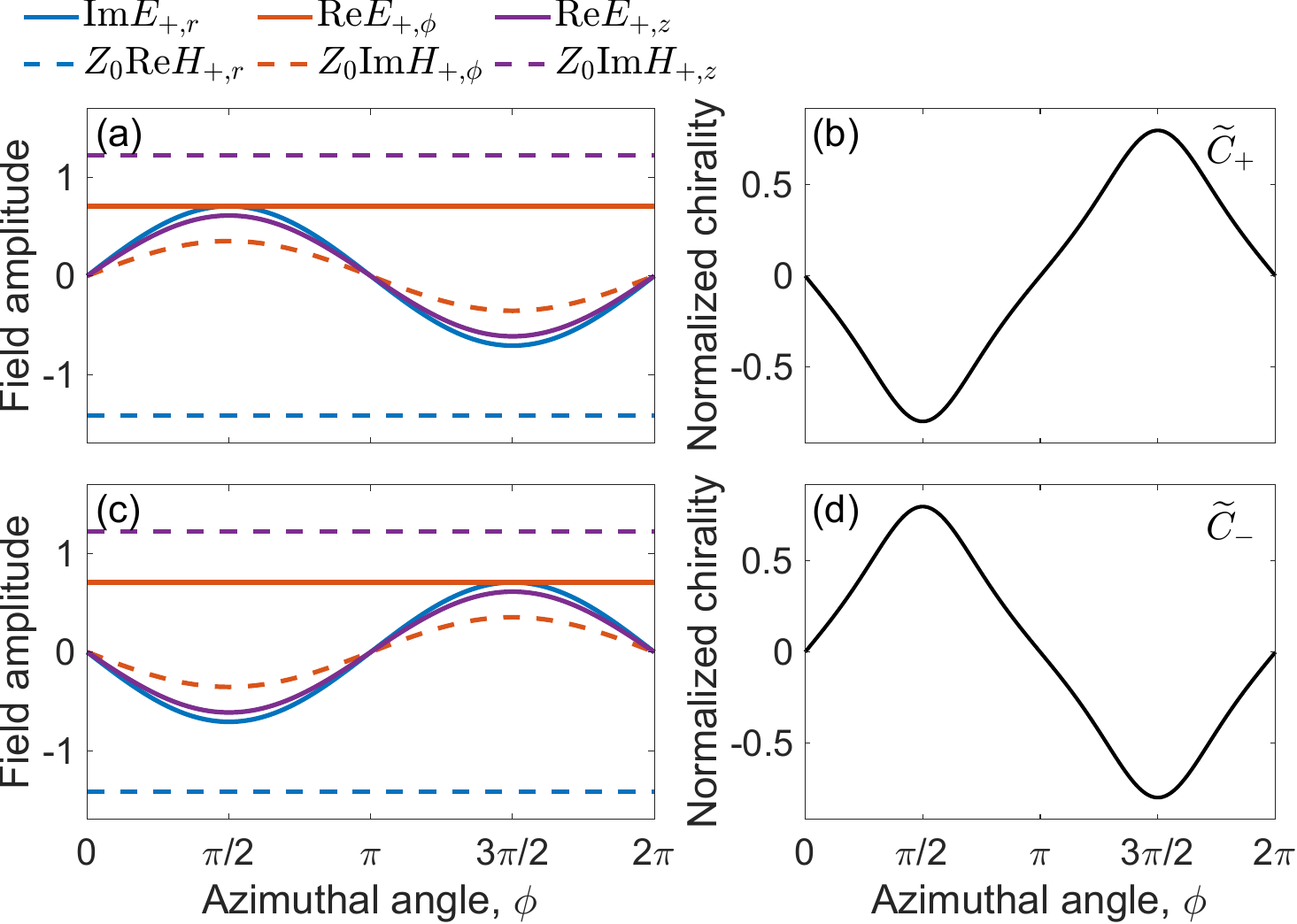}}
    \caption{Azimuthal dependence of the field components and optical chirality in the cladding of a cylindrical dielectric waveguide. The upper and lower rows correspond to the superpositions of the TE$_{01}$ and HE$_{11}^{y}$ modes defined by Eq.~\eqref{eq:superpositions} with $\alpha=\pm i$, respectively. Panels (a,c) show the real and imaginary field components contributing to $\operatorname{Im}(\mathbf E\cdot\mathbf H^*)$, as indicated in the legend. Solid and dashed curves represent electric and magnetic fields, respectively; blue, orange, and purple denote the $r$, $\phi$, and $z$ components. Electric fields and impedance-scaled magnetic fields $Z_0\mathbf H$ are normalized by the common radial amplitude. Panels (b,d) show the normalized optical chirality $\widetilde C_{\pm}$ according to Eq.~\eqref{eq:C_norm}. All plotted quantities are independent of the radial coordinate.}
	\label{fig:cyl_WG}
\end{figure}

Direct substitution of Eqs.~\eqref{eq:TE_clad} and \eqref{eq:HE_clad} into Eq.~\eqref{eq:C_def} yields
\begin{equation}
C_{\mathrm{TE}}(\mathbf r)=C_{\mathrm{HE}}(\mathbf r)=0.
\label{eq:TE_HE_C}
\end{equation}
Thus, both individual modes exhibit zero local optical chirality within this approximation. Independently of this local result, mirror symmetry requires the chirality integrated over the complete cross-section to vanish for either mode.

Nonzero local chirality can nevertheless arise from a coherent superposition of the two modes. For nondegenerate modes, however, their relative phase remains nearly constant only over distances much shorter than the characteristic length $L_{\mathrm{ch}}\sim 1/|\beta_1-\beta_2|$. We therefore first consider the ideal case of degenerate modes, $\beta_1=\beta_2=\beta$, for which the superposition retains its transverse field profile at any propagation distance, apart from a common propagation factor. Within the present cladding approximation, this also gives $\gamma_1=\gamma_2=\gamma$ and $F_1(r)=F_2(r)=F(r)$.

To determine the relative amplitude and phase that maximize optical chirality, we consider a general superposition of the two modes:
\begin{equation}
\mathbf E_\alpha=\mathbf E_{\mathrm{TE}}+\alpha\mathbf E_{\mathrm{HE}},
\qquad
\mathbf H_\alpha=\mathbf H_{\mathrm{TE}}+\alpha\mathbf H_{\mathrm{HE}},
\label{eq:superpositions}
\end{equation}
where $\alpha$ is a complex relative amplitude. We choose the modal normalization such that the azimuthal electric-field amplitudes at $r=R$ are equal. With this convention, $a_1=a_2=1$, and the relative excitation amplitude and phase are contained entirely in $\alpha$. Because the self-chirality of each mode vanishes, the chirality of the superposition is determined entirely by the interference terms between the TE$_{01}$ and HE$_{11}^y$ fields:
\begin{equation}
C_s(r,\phi)
=
-\varepsilon_0\beta |F(r)|^2\sin\phi\,\operatorname{Im}(\alpha).
\label{eq:chirality_distribution}
\end{equation}

Equation~\eqref{eq:chirality_distribution} shows that nonzero chirality requires a nonzero relative phase between the modes and changes sign across the symmetry plane $y=0$. Consequently, the chirality integrated over the complete cross-section vanishes for any $\alpha$. At the same time, the chirality integrated over either half of the cladding is nonzero when $\operatorname{Im}(\alpha)\neq0$. For example, as illustrated in Figs.~\ref{fig:cyl_WG}(a,c), for $\alpha=\pm i$ the same components of electric and magnetic fields acquire $\pm \pi/2$ phase difference. Then, in the upper (or lower) half-area the integrated chirality becomes non-zero,
\begin{equation}
C_s^{(\mathrm{up})} = \int_{S_{\mathrm{up}}}C_s(\mathbf r)\,dS \ne 0,
\end{equation}
where $S_{\mathrm{up}}: \; 0\leq\phi\leq\pi,\; R\leq r<\infty$.

To introduce a quantitative metric, we define the normalized optical chirality as
\begin{equation}
\widetilde C(\mathbf r)
=
\frac{C(\mathbf r)}{k_0W(\mathbf r)},
\label{eq:C_norm}
\end{equation}
where
\begin{equation}
W(\mathbf r)
=
\frac{1}{4}
\left[
\varepsilon_0\varepsilon(\mathbf r)|\mathbf E(\mathbf r)|^2
+
\mu_0|\mathbf H(\mathbf r)|^2
\right]
\label{eq:Wav}
\end{equation}
is the electromagnetic energy density in a dispersionless, nonmagnetic medium. For a circularly-polarized plane wave in vacuum $\widetilde C(\mathbf r)$ achieves extreme values 
$\pm 1$ everywhere in space~\cite{HanifehACS2020}. For the waveguide considered here, the normalized chirality reaches values up to $\pm0.7$, as shown in Figs.~\ref{fig:cyl_WG}(b,d). The normalized chirality averaged over an area or volume $\Omega$ is then defined as
\begin{equation}
\widetilde C^{\mathrm{av}}_{\Omega}
=
\frac{\displaystyle\int_\Omega C(\mathbf r)\,d\Omega}
{k_0\displaystyle\int_\Omega W(\mathbf r)\,d\Omega},
\label{eq:C_av_norm}
\end{equation}
where $d\Omega=dS$ for an area average and $d\Omega=dV$ for a volume average.

Under the adopted modal normalization, the magnitude of the chirality averaged over the upper half-area is maximized for
\begin{equation}
\alpha
=
\pm i\sqrt{\frac{2\widetilde{\beta}^2}
{1+\widetilde{\beta}^2}},
\end{equation}
and the maximum value is
\begin{equation}
\max_{\alpha}
\left|\widetilde C^{\mathrm{av}}_{\mathrm{up}}\right|
=
\frac{2\sqrt{2}}
{\pi\sqrt{1+\widetilde{\beta}^2}}.
\label{eq:C_up_norm}
\end{equation}
Thus, the maximum is obtained when the modes have a relative phase of $\pm\pi/2$ and comparable amplitudes. For a representative effective index $\widetilde{\beta}\approx1.5$, the maximum upper-half-space-averaged chirality is approximately $0.5$ of the circular-plane-wave reference value. Therefore, although the chirality integrated over the complete cross-section vanishes, a molecular layer placed on one side of the waveguide can experience substantial single-signed optical chirality. This suggest a route to achieve measurable optical activity and circular dichroism over an extended waveguide interaction length with considered modes, provided that their propagation constants are sufficiently close to each other.

\section{Design of the periodic waveguide}

The required near degeneracy condition of the guided modes can be realized through dispersion engineering of subwavelength-grating waveguides~\cite{HalirLPR2015,ChebenNature2018}, as demonstrated previously for related dielectric waveguide systems~\cite{SavelevJAP2019,MikhinNL2023}. For chiral sensing, the operating spectral range is determined by the relevant molecular transitions: electronic circular dichroism is typically measured in the ultraviolet and visible ranges, whereas VCD probes vibrational transitions in the mid-infrared range~\cite{KiederlingMolecules2018,WarningACS2021}. Possible mid-infrared implementations include silicon or germanium structures on low-loss infrared-transparent substrates~\cite{YeACS2023}, as well as suspended silicon and germanium waveguide platforms~\cite{PenadesOL2018,OsmanOL2018,LiuNanophotonics2021,PostigoOE2021}. In the visible range, gallium phosphide could potentially be employed. At shorter wavelengths, diamond nanostructures can support resonances suitable for ultraviolet chiral spectroscopy~\cite{HuACSPhotonics2020}, whereas extension into the deep-ultraviolet range remains challenging because of the scarcity of high-index, low-loss materials~\cite{ShenACS2025}.

\begin{figure}[t]
    \center{\includegraphics[width=1\columnwidth]{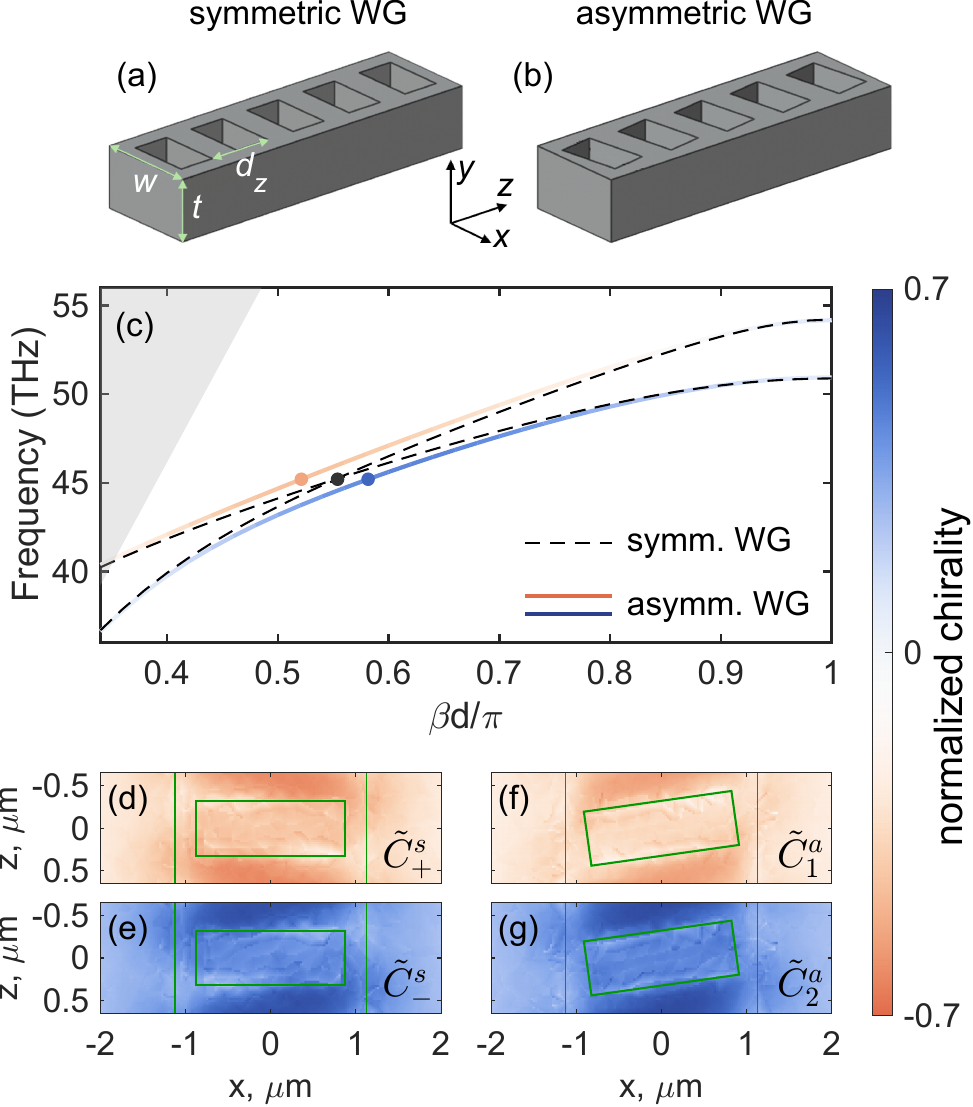}}
	\caption{(a,b) Schematics of the suspended SWG waveguide with (a) a mirror-symmetric unit cell and (b) a symmetry-broken unit cell containing a rotated rectangular hole. (c) Dispersions of the symmetric waveguide (black dashed curves) and the asymmetric waveguide (colored solid curves). The gray shaded region denotes the radiation continuum above the light line. The color represents the normalized chirality $\widetilde{C}_{\mathrm{av}}$ averaged over the upper half-space. The three solid circles indicate the common frequency used in panels (d--g): the degeneracy point of the two symmetric-waveguide modes and the two asymmetric-waveguide modes at different Bloch wave numbers. (d-g) Spatial distributions of the normalized chirality density $\widetilde{C}(\mathbf r)$ in a plane located 25~nm above the upper waveguide surface for (d,e) the symmetric-waveguide superpositions and (f,g) modes of the asymmetric waveguide, calculated at frequency of degeneracy $\approx 45$~THz. Green contours indicate the germanium boundaries. The same color scale is used for $\widetilde{C}_{\mathrm{av}}$ in (c) and for the $\widetilde{C}(\mathbf r)$ in (d--g). }
	\label{fig:GeWG}
\end{figure}

In this work, we restrict our consideration to waveguides operating in the mid-infrared spectral range. The designed two-mode suspended germanium waveguide is shown schematically in Fig.~\ref{fig:GeWG}(a). Its parameters are as follows: waveguide width $w=2.25$~$\mu$m, membrane thickness $t=1.4$~$\mu$m, longitudinal period $d_z=1.3$~$\mu$m, and hole dimensions $h_x \times h_z=1.75$~$\mu$m $\times$ $0.64$~$\mu$m. The germanium refractive index is assumed to be real, frequency independent, and equal to $n=4$. The dispersions of the two modes of interest are shown by the black dashed curves in Fig.~\ref{fig:GeWG}(c). The modes have the same parity with respect to the horizontal midplane of the membrane and are qualitatively similar to the TE and HE modes of the cylindrical waveguide considered in the previous section. Dispersion engineering brings the two modes to degeneracy at a frequency of approximately 45~THz.

The normalized chirality density $\tilde{C}_{\pm}^s$ [Eq.~\eqref{eq:C_norm}] of the two superpositions, Eq.~\eqref{eq:superpositions} with $\alpha=\pm i$, 
evaluated at the degeneracy point, is shown in Figs.~\ref{fig:GeWG}(d,e). Before constructing the superpositions, the fields of the two modes were normalized to equal power. The distributions are evaluated in a plane located 25~nm above the upper waveguide surface. They are relatively uniform above the waveguide and have opposite signs for the two opposite phase relations. The average normalized chirality [Eq.~\eqref{eq:C_av_norm}] is $\widetilde{C}_{\mathrm{av}}\approx\pm0.4$ ($\pm0.5$) for a 100~nm (500~nm) thick layer above the waveguide. The averaging is performed over one longitudinal period and the entire transverse extent of the layer.

In the symmetric waveguide [Fig.~\ref{fig:GeWG}(a)], the chiral field therefore relies on the coherent superposition of two degenerate or nearly degenerate modes. Consequently, the chirality remains close to its initial value only over propagation distances of the order of $L_{\mathrm{ch}}\sim 1/|\beta_1-\beta_2|$, after which modal beating substantially modifies the chirality distribution~\cite{VazquezLozano2019}. Thus, for propagation distances much larger than the unit-cell period, the symmetric waveguide operates as intended only within a narrow frequency range around the modal degeneracy.

This restriction can be relaxed by introducing a geometrical perturbation that couples the two modes in such a way that the hybridized Bloch eigenmodes themselves carry nonzero chirality in the upper half-space. Importantly, the chirality integrated over the entire waveguide cross-section remains zero. It is therefore unnecessary to eliminate all mirror planes through a three-dimensionally chiral modification, such as twisting the waveguide~\cite{HelfrichJPP2026}; breaking the relevant in-plane symmetry of the unit cell, as illustrated in Fig.~\ref{fig:GeWG}(b), is sufficient. Therefore, unlike the symmetric waveguide, a prescribed phase-locked superposition of two independently excited modes is no longer required.

The dispersions of the hybridized modes are shown by the colored solid curves in Fig.~\ref{fig:GeWG}(c), while their chirality distributions are presented in Figs.~\ref{fig:GeWG}(f,g). The latter are evaluated at the same frequency as the symmetric-waveguide degeneracy, but at the two different Bloch wave numbers indicated in Fig.~\ref{fig:GeWG}(c). The local chirality distributions of the hybridized modes very closely reproduce those of the symmetric-waveguide superpositions, retaining a predominantly uniform sign above the waveguide. Moreover, the branch colors in Fig.~\ref{fig:GeWG}(c) show that the average chirality remains of the same order as that of the symmetric-waveguide superpositions over a spectral range of approximately 10\% of the central frequency.

\section{Perturbation theory for a thin chiral layer}

To analyze the specific characteristics of chirality-induced effects, we next consider a periodic waveguide covered by a thin chiral layer of thickness $t_{ch}$ occupying the region above the waveguide [see inset in Fig.~\ref{fig:CD_OR_all}(a)] and characterized by the Pasteur constitutive relations:
\begin{equation}
\begin{aligned}
\VB{D}(\VB{r}) &= \eps_0\eps(\VB{r}) \VB{E}(\VB{r}) + i \chi(\VB{r})/c \VB{H}(\VB{r}),
\\
\VB{B}(\VB{r}) &= \mu_0\mu(\VB{r}) \VB{H}(\VB{r}) - i \chi(\VB{r})/c \VB{E}(\VB{r}).
\end{aligned}
\label{Pasteur}
\end{equation}
Here, $\eps$, $\mu$, and $\chi$ are the permittivity, permeability, and Pasteur parameter, respectively. In what follows, as a representative example we consider the following parameters of the chiral layer: ${t_{ch}=500}$~nm, ${\eps_{ch}=2.25 + 3\cdot10^{-3}i}$, ${\mu_{ch}=1+3\cdot10^{-3}i}$, and ${\chi=(1+i)\cdot10^{-6}}$, ensuring the passivity of the chiral medium~\cite{SihvolaMOTL2001}.

For small perturbations, the propagation constants $\beta_c$ of the coupled modes in the presence of the chiral layer can be derived within the framework of perturbation theory from the following matrix equation:
\begin{equation}
\begin{pmatrix}
\tilde{\beta}_c - \tilde{\beta}_1 - V_{11} & -V_{12} \\
-V_{21} & \tilde{\beta}_c - \tilde{\beta}_2 - V_{22}
\end{pmatrix}
\begin{pmatrix}
c_1 \\
c_2
\end{pmatrix}
=0,
\label{eq:Matrix_equation}
\end{equation}
where $\tilde{\beta}_{1,2,c} = \beta_{1,2,c}/k_0$, and the coupling constants $V_{mn}$ are calculated as the following overlap integrals over the volume of the chiral layer ($V_{ch}$):
\begin{equation}
\begin{aligned}
V_{mn}
=
\dfrac{1}{4P_m d}
\int \limits_{V_{ch}}
\left[
\dfrac{1}{Z_0} \Delta\eps \VB{e}_m^{(0)*} \cdot \VB{e}_n^{(0)}
+
Z_0 \Delta\mu \VB{h}_m^{(0)*} \cdot \VB{h}_n^{(0)} + \right.\\
+ \left. i\chi(\VB{e}_m^{(0)*} \cdot \VB{h}_n^{(0)} - \VB{e}_n^{(0)} \cdot \VB{h}_m^{(0)*})
\right]
\, dV,
\end{aligned}
\label{eq:coupling_constants}
\end{equation}
where $P_m = \dfrac{1}{2} \Re \int \limits_{CS} \left(\VB{e}_m^{(0)} \times{\VB{h}_m^{(0)}}^*\right)\cdot\hat{\mathbf{z}}\,dS$, $m=1,2$, is the power flow carried by the $m$th mode through an arbitrary transverse plane, and $\VB{e}_m^{(0)},\VB{h}_m^{(0)}$ are the Bloch wavefunctions of the unperturbed modes. Furthermore, $\Delta\eps = \eps_{ch} - \eps_{wg}$ and $\Delta\mu = \mu_{ch} - \mu_{wg}$, where $\eps_{wg}$ and $\mu_{wg}$ are the permittivity and permeability, respectively, of the unperturbed structure in the volume occupied by the chiral layer. For convenience, in further calculations the real part of the chiral-layer permittivity is incorporated into the unperturbed waveguide by setting $\eps_{wg}=2.25$. Consequently, all fields and coupling coefficients are calculated for the waveguide covered with the layer with real permittivity. Then, the small imaginary-valued perturbations $\Delta\eps$ and $\Delta\mu$ only induce additional losses in the waveguide modes. See \purple{Supporting Information, Sec. II} for the further details on the perturbation theory.

Owing to the symmetries of the modes and the perturbation, the last term in Eq.~\eqref{eq:coupling_constants} vanishes for $V_{nn}$, indicating zero integral chirality of individual modes, (as elaborated in Supporting Information, Sec. III). Whereas the first two terms, which are proportional to the imaginary-valued perturbations $\Delta\eps$ and $\Delta\mu$, only induce additional imaginary parts of the propagation constants. Consequently, one can redefine the modal dispersions as
\begin{equation}
\tilde{\beta}^{(corr)}_{n} = \tilde{\beta}_{n} + V_{nn} \equiv \tilde{\beta}_{n} + i \gamma^{ch}_n,
\end{equation}
where
\begin{equation}
\gamma^{ch}_n = \dfrac{1}{\tilde{v}_{g_n}}(\eta_n^{e}\dfrac{\eps_{ch}''}{\eps_{ch}'} + \eta_n^{m}\mu_{ch}''),
\label{eq:V_nn_corr}
\end{equation}
where $\tilde{v}_{g_n}$ is the group velocity of the $n$th mode normalized by the speed of light, ${\eta_n^{e} = \int \limits_{V_{ch}} W^e_n dV/\int \limits_{V_{UC}} W_n dV }$, ${\eta_n^{m} = \int \limits_{V_{ch}} W^m_n dV/\int \limits_{V_{UC}} W_n dV}$, and $W^e_n$, $W^m_n$, $W_n$ are electric, magnetic and total electromagnetic energy density, respectively, of the $n$th mode of the lossless waveguide, Eq.~\eqref{eq:Wav}; $V_{ch}$ is the volume of the chiral layer, and $V_{uc}$ is the volume of the whole unit cell. Here and throughout the paper, $'$ and $''$ denote the real and imaginary parts, respectively. The fields of the modes were normalized to unit power. Corrected propagation constants $\tilde{\beta}^{(corr)}_n$ are now complex-valued even for an initially lossless waveguide ($\tilde{\beta}_n''=0$). Note that if $\eps_{ch}'' \gg \mu_{ch}''$, the electric-energy overlap $\eta_n^{e}$ becomes particularly important~\cite{Tang2010}. However, waveguide modes whose superposition produces chiral fields typically require similar electric and magnetic energy overlaps, $\eta_n^{e} \approx \eta_m^{e}$.


For the same symmetry reasons, only the third term in Eq.~\eqref{eq:coupling_constants} is nonzero for the cross-coupling coefficients $V_{12}$ and $V_{21}$. Moreover, owing to the $z=0$ symmetry plane, the corresponding overlap integral containing $(\VB{e}_m^{(0)*} \cdot \VB{h}_n^{(0)} - \VB{e}_n^{(0)} \cdot \VB{h}_m^{(0)*})$ can be made real-valued by a proper choice of the relative global phase of the two waveguide modes \purple{(see Supporting Information, Sec. III)}. The coefficients $V_{12}$ and $V_{21}$ can then be expressed through the characteristics of the fields formed by the superposition of the two waveguide modes, $\mathbf E_{\pm} = \VB{E}_1 \pm i\VB{E}_2$ and $\mathbf H_{\pm} = \VB{H}_1 \pm i\VB{H}_2$, where the fields of the uncoupled modes are normalized to unit power:
\begin{equation}
V_{12,21} =  -i \chi \dfrac{\eta}{\tilde{v}_g^{av}} \widetilde{C}^{av}_{\pm}\equiv \pm i\delta ,
\label{eq:coupling_constants_simplified}
\end{equation}
where the $\pm$ subscript denotes the two superpositions of the modes, $\eta=\int_{V_{ch}} W_{\pm}(\VB{r}) \,dV/\int_{V_{uc}} W_{\pm}(\VB{r}) \,dV$ is the fraction of electromagnetic energy in the region $V_{ch}$ relative to that in the entire unit-cell volume $V_{uc}$, and $\tilde{v}_g^{av}$ is the harmonic mean of the normalized group velocities $\tilde{v}_{g_{1,2}}$. The quantity $\widetilde{C}^{av}_{\pm}$, defined according to Eq.~\eqref{eq:C_av_norm}, is averaged over $V_{ch}$. Note that due to opposite symmetry of two modes with respect to $x=0$ plane $\widetilde{C}^{av}_{+}=-\widetilde{C}^{av}_{-}$.
The resulting coupled-mode equation reads:
\begin{equation}
\det
\begin{pmatrix}
\tilde{\beta}_c - \tilde{\beta}^{(corr)}_1 & - i\delta \\
i\delta & \tilde{\beta}_c - \tilde{\beta}^{(corr)}_2
\end{pmatrix}
= 0.
\label{eq:CMT_symm_final}
\end{equation}

Using the described framework, we analyze two distinct cases of a waveguide covered by a thin chiral layer: (i) the symmetric waveguide supporting two quasi-degenerate modes, Fig.~\ref{fig:GeWG}(a), and (ii) the asymmetric waveguide with rotated holes supporting two non-degenerate modes characterized by nonzero chirality above the waveguide, Fig.~\ref{fig:GeWG}(b). In the latter case, the geometrical coupling of the waveguide modes induced by the rotation of the holes can be described by the same coupled-mode matrix equation by introducing an additional geometrical perturbation, leading to extra non-diagonal term $\delta_{rot}$: $\delta_{as} = \delta + \delta_{rot}$. It can be shown that $\delta_{rot}$ is real-valued owing to symmetry considerations (\purple{see Supporting Information, Sec. III}), and its sign corresponds to the direction of the holes rotation.

First, we calculate the waveguide dispersions using Eqs.~\eqref{eq:V_nn_corr}, \eqref{eq:coupling_constants_simplified}, and \eqref{eq:CMT_symm_final}, as well as by direct numerical simulations in COMSOL Multiphysics for comparison~\cite{EnriquezJOSAb2023,LeeSciRep2018}. Next, we use the obtained dispersions to calculate two measurable transmission characteristics of a waveguide covered by the chiral layer and having a normalized length $L=k_0l$, where $l$ is the physical propagation distance, assuming that it is excited from the corresponding achirally coated reference waveguide. The first characteristic is the differential transmission $\Delta T$, which is analogous to circular dichroism when reflectance is neglected. The second characteristic is the modal power conversion, which is analogous to optical rotation.

For the symmetric waveguide, $\Delta T$ is calculated by launching combinations of the TE and HE modes with opposite $\pm \pi/2$ phase differences and appropriate amplitudes into the waveguide covered by the chiral layer. The difference between the total output powers, normalized by the total input power, is then calculated:
\begin{equation}
\Delta T_s = \dfrac{P^{+}_{out} - P^{-}_{out}}{P_{in}},
\end{equation}
where the $\pm$ indices correspond to the opposite phase differences between the modes at the input. Modal conversion is calculated by launching one of the modes, TE or HE, at the input and determining the normalized power of the opposite mode at the output:
\begin{equation}
R_s = \dfrac{P^{m}_{out}}{P^{n}_{in}},
\end{equation}
where $m,n=1,2$ or $2,1$.

For a single asymmetric waveguide, the differential transmission can be determined by calculating the difference between the output powers $P^{a_j}_{out}$, $j=1,2$, corresponding to the excitation of either of the two coupled modes:
\begin{equation}
\Delta T_a = \dfrac{P^{a_1}_{out} - P^{a_2}_{out}}{P_{in}}.
\end{equation}
This characteristic is suitable close to the mode-degeneracy point, where the fields of two modes are approximately the mirror counterparts. At large spectral detunings from the degeneracy point, however, such relation fails and the losses associated with the two branches generally become different, which compromises the purity of the differential sensing signal~\cite{GrafACSPhotonics2019}.

The alternative way to define the differential transmission that avoids this limitation is to consider two waveguides with opposite directions of hole rotation. This configuration is straightforward to implement because it can be fabricated on the same substrate and requires only an in-plane modification of the waveguide geometry. The same corresponding mode $a_j$ is then excited in both waveguides, and the differential transmission is defined as
\begin{equation}
\Delta T_r^{(j)} =
\dfrac{P^{(+\theta,a_j)}_{out} - P^{(-\theta,a_j)}_{out}}{P_{in}},
\qquad j=1,2,
\label{eq:Delta_T_rotation}
\end{equation}
where $+\theta$ and $-\theta$ denote the opposite directions of hole rotation. Since the two uncovered waveguides are related by an in-plane mirror transformation, the corresponding achiral reference modes are exact mirror counterparts over a broad spectral range. Therefore, $\Delta T_r^{(j)}$ provides a cleaner differential sensing signal than $\Delta T_a$ away from the mode-degeneracy point.

We do not consider power conversion between the two uncoupled modes in the asymmetric waveguide because geometrical coupling leads to a very strong dependence of the converted power on the propagation distance, thereby obscuring the modal conversion induced by the chiral perturbation.

\section{Transmission characteristics}

To obtain transmission characteristics in the spectral range near the degeneracy point we solve Eq.~\eqref{eq:CMT_symm_final}, assuming the following dispersion of diagonal elements: $\tilde{\beta}^{(corr)}_{n} \approx \tilde{\beta}_0(1-\Delta\tilde{\omega}) + \Delta\tilde{ \omega}/\tilde{v}_{gn} + i\tilde{\beta}_{n}'' + i\gamma^{ch}_n$, $n=1,2$, where $\tilde{\beta}_0=1.4$ is the wavenumber corresponding to the degeneracy point, $\Delta\tilde{ \omega} = \Delta \omega/\omega_0$ is the relative spectral detuning, $\tilde{\beta}_{n}''$ are waveguide modes losses assumed to be related mainly to the surface roughness. We neglect the frequency dependence of the model parameters over the considered spectral range and use their values calculated at the degeneracy point: $\tilde{v}_{g1}=0.164$, $\tilde{v}_{g2}=0.23$, $\gamma^{ch}_1 = 2.66\cdot10^{-4}$, $\gamma^{ch}_2 = 2.84\cdot10^{-4}$, $\eta = 0.024$, $|\widetilde{C}^{av}_{\pm}| = 0.42$, $\tilde{v}_g^{av} = 0.191$, which yields $\delta=0.5\cdot(1+i)\cdot10^{-7}$. 

The total modal dichroism $\Delta \beta'' = \Delta \tilde\beta'' + \Delta\gamma^{ch} $, where $\Delta \beta = \tilde{\beta}^{(corr)}_1 - \tilde{\beta}^{(corr)}_2$, consists of bare waveguide contribution $\Delta \tilde\beta'' = \tilde{\beta}_{1}''-\tilde{\beta}_{2}''$ and contribution from chiral layer $\Delta\gamma^{ch} = \gamma^{ch}_1-\gamma^{ch}_2$ due to achiral losses $\eps''$ and $\mu''$. The modal dichroism induced by the chiral layer for the chosen parameters $\eps''=\mu''=3\cdot10^{-3}$ is relatively weak: $|\Delta\gamma^{ch}| \approx 2\times10^{-5}$. The rough assumption of the bare waveguide modal dichroism $|\Delta \tilde\beta''|\lesssim 10^{-4}$ can be made based on the losses in the considered suspended waveguides reported in the literature~\cite{PenadesOL2018,LiuNanophotonics2021,PostigoOE2021}, which are up to a few dB/cm for the fundamental mode. Transmission characteristics are then calculated for three values of the total modal dichroism ${\Delta\beta''=-\{10^{-6},\; 10^{-5},\; 10^{-4}\} }$ and fixed average losses $\beta_s''=3.3\cdot10^{-4}$, where $\beta_s=(\tilde{\beta}^{(corr)}_1+\tilde{\beta}^{(corr)}_2)/2$. In the calculations, the total losses of two modes were taken as $\beta_s'' \pm \Delta\beta''/2$.

In the following calculations we choose the normalized length of the waveguide $L=1500$ ($l\approx 1.5$~mm at the given wavelength), which approximately corresponds to the maximal differential transmission at degeneracy frequency, as will be shown further. For the chosen length of the waveguide one can safely assume that accumulated chiral coupling is weak, $|\delta|L \approx 10^{-4} \ll 1$, which is used further when deriving simplified expressions for transmission characteristics.

\begin{figure}[t]
\center{\includegraphics[width=1\columnwidth]{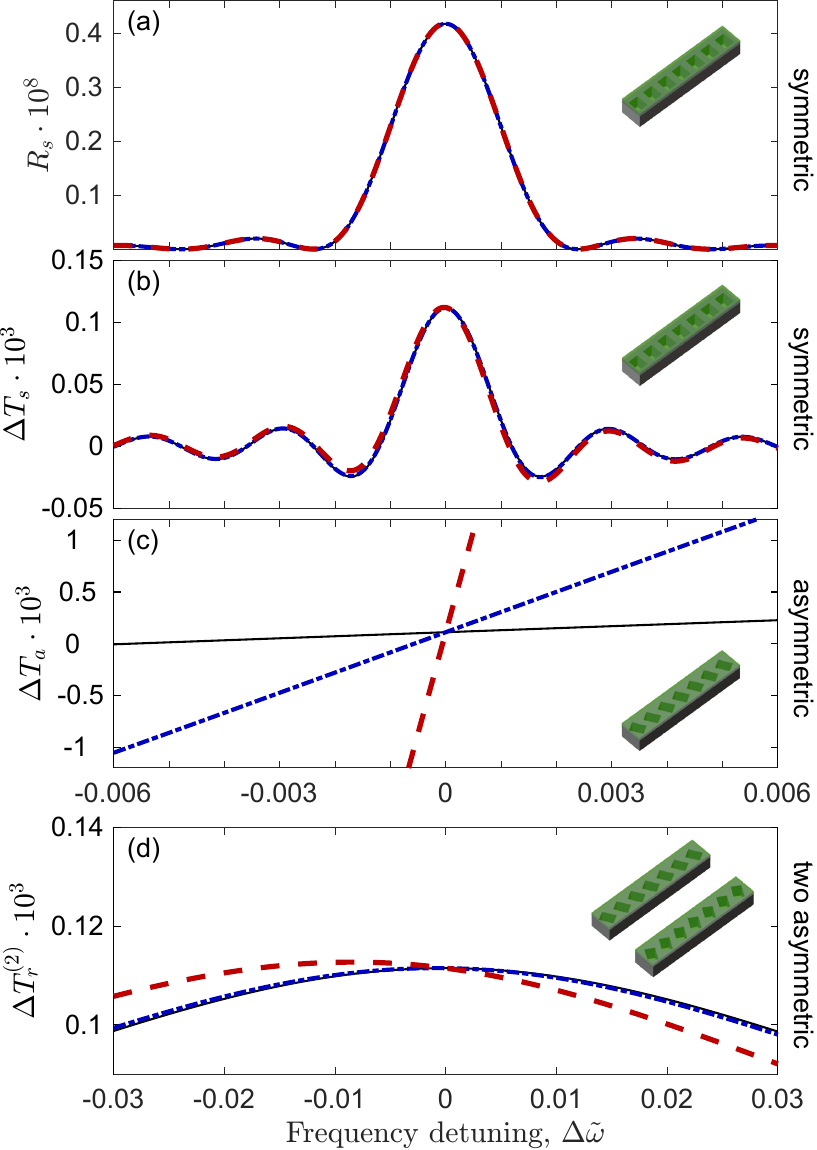}}
	\caption{Transmission characteristics calculated with exact expressions from the coupled-mode theory for three values of modal dichroism $\Delta\beta''=-10^{-6}$ (black solid curves), $-10^{-5}$ (blue dash-dotted curves), and $-10^{-4}$ (red dashed curves). (a) Modal power conversion $R_s$ and (b) differential transmission $\Delta T_s$ for the symmetric waveguide. (c) Differential transmission $\Delta T_a$ for a single asymmetric waveguide. (d) Rotation-differential transmission $\Delta T_r^{(2)}$ for two asymmetric waveguides with opposite hole-rotation angles. The insets schematically show the corresponding waveguide configurations. Note that panels (a-c) and (d) use different horizontal scales.}
	\label{fig:CD_OR_all}
\end{figure}

\textbf{Symmetric waveguide.} Spectral dependence of $R_s$ and $\Delta T_s$ are plotted in Figs.~\ref{fig:CD_OR_all}(a,b) using the exact solutions (see Supporting Information, Sec. IV for the details). The calculated dependencies near the degeneracy frequency can be understood within the approximation $|\Delta\beta|L \ll 1$. In this case transmission characteristics can be written as follows: 
\begin{equation}
\begin{aligned}
R_s &\approx
e^{-2\beta_s''L}|\delta|^2L^2
\left[
1+
\dfrac{\left(\Delta\beta''^2-\Delta\beta'^2\right)L^2}{12}
\right],\\
\Delta T_s &\approx
4e^{-2\beta_s''L}L
\left[
\delta'' + \right.\\
& \left. + \dfrac{L^2}{6}
\left(
\delta'\Delta\beta'\Delta\beta''
+
\delta''
\left[\Delta\beta''^2-\Delta\beta'^2\right]
\right)
\right].
\end{aligned}
\label{eq:Rs_DeltaTs_small_detuning}
\end{equation}

One can observe that $R_s$ always exhibits maximum at the degeneracy point $\Delta\beta'=0$ regardless of modal dichroism. It is proportional to the $|\delta|^2$ (or $|\chi|^2$), since we are considering the intensity redistribution between the modes and not the polarization rotation (amplitude characteristic) as in conventional optical rotation experiments. Consequently, very small values of $R_s$ of the order of $10^{-8}$ make this characteristic of limited use in practical systems. Note also that intensity measurement allows only for determining the absolute value of Pasteur parameter $|\chi|$, while distinction between the left and right chirality requires phase-sensitive measurement of the converted mode. Near the degeneracy point, the leading reduction of $R_s$ is proportional to $\Delta\beta'^2$, reducing significantly for spectral detuning $|\Delta\beta'| \gtrsim 1/L$. 

On the other hand, near the degeneracy frequency $\Delta T_s$ is linearly proportional to $\delta''$, leading to the potentially measurable maximum value $\Delta T_s^{\mathrm{max}} \approx 4e^{-2\beta_s''L}
\delta''L \approx 1.1\cdot10^{-4}$. At the same time $\Delta T_s$ contains a term proportional to $\Delta\beta'\Delta\beta''\delta'$, producing an asymmetric, locally linear spectral contribution and mixing the real and imaginary parts of the chiral coupling~\cite{DrouliasPRB2020}. This term shifts the maximum of $\Delta T_s$ from the degeneracy point by the value $\propto \left|\dfrac{\delta'}{\delta''}\Delta\beta''\right|$. As soon as $|\Delta\beta''| L \ll \left|\dfrac{\delta''}{\delta'}\right|$ this shift is negligible. 
For the parameters in Fig.~\ref{fig:CD_OR_all}(b) this condition is approximately satisfied, 
and $\Delta T_s$ also exhibits parabolic spectral response near the degeneracy frequency. Consequently, symmetric waveguide allows extraction of $\chi''$ only in the narrow spectral range limited by $|\Delta\beta'| \lesssim 1/L$.

\textbf{Asymmetric waveguide.} The operational spectral range can be substantially expanded by considering an asymmetric waveguide with rotated holes [see inset in Fig.~\ref{fig:CD_OR_all}(c)]. The additional non-diagonal term $\delta_{\mathrm{rot}}$ due to rotation of the holes by 8 degrees was estimated from the numerically calculated dispersion splitting as $\delta_{\mathrm{rot}}\approx 0.05$, which is several orders larger than $|\delta|$. To capture the main features of the transmission spectrum near the degeneracy frequency, we have expanded $\Delta T_a$ to the second order in $|\Delta \beta|/\delta_{\mathrm{rot}}$ (as well as to first order in $\delta/\delta_{\mathrm{rot}}$) to obtain \purple{(see Supporting Information, Sec. V for the details)}:
\begin{equation}
\Delta T_a
    \approx
    -e^{-2\beta_s''L}\left(\dfrac{\Delta\beta'\Delta\beta''}{|\delta_{\mathrm{rot}}|} + 4\dfrac{\delta_{\mathrm{rot}}}{{|\delta_{\mathrm{rot}}|}}
        \delta''
        \right) \zeta L,
\label{eq:DeltaT_a}
\end{equation}
where $\zeta = \left( 1 - \dfrac{\Delta\beta'^2}{8\delta_{\mathrm{rot}}^2} \right)$. Although this approximation is formally valid only for $|\Delta \beta| \ll \delta_{\mathrm{rot}}$, it captures the main trend over a broader range, approximately up to $\Delta\beta' \lesssim \delta_{\mathrm{rot}}$ (assuming that modal dichroism is considerably small $|\Delta \beta''| \ll \delta_{\mathrm{rot}}$).

One can observe, that besides the term proportional only to the $\delta''$, $\Delta T_a$ contains an additional background term that does not depend on the chirality parameter $\delta$, but is linearly proportional to the spectral detuning $\Delta \beta'$, modal dichroism $\Delta \beta''$, and is independent of the direction of the holes rotation (sign of the $\delta_{\mathrm{rot}}$). This is illustrated in Fig.~\ref{fig:CD_OR_all}(c). Consequently, in the case of strong modal dichroism the spectral range where the background is smaller than the useful chiral signal becomes limited by $\Delta \beta' \lesssim \left| 4\dfrac{\delta_{\mathrm{rot}}\delta''}{\Delta \beta''} \right|$. For the maximal considered modal dichroism such constraint $\Delta \beta' \lesssim 10^{-4}$ is even stronger than $|\Delta\beta'| \lesssim 6.7\cdot10^{-4}$ limit in the symmetric waveguide.

This unwanted achiral background can be removed in the corresponding achiral system composed of two uncoupled waveguide with opposite structural asymmetry. In this case rotation-differential transmission spectra read:
\begin{equation}
\Delta T_r^{(1,2)} \approx \mp 4 e^{-2\beta_s''L}\dfrac{\delta_{\mathrm{rot}}}{{|\delta_{\mathrm{rot}}|}}\delta''L\, \left( 1 \mp \dfrac{\Delta\beta'\Delta\beta''}{2|\delta_{\mathrm{rot}}|}L \right)\zeta.
\label{eq:DeltaT_r}
\end{equation}
The spectrum of $\Delta T_r^{(j)}$ still contains additional asymmetric term $\dfrac{\Delta\beta'\Delta\beta''}{2|\delta_{\mathrm{rot}}|}L$. However, even for the largest considered $\Delta\beta''$ the magnitude of the asymmetry is $1.5 |\Delta \beta'|$, which can be neglected since we are working in the spectral range $\Delta\beta' \lesssim \delta_\mathrm{rot}$. Thus the $\Delta T_r^{(1,2)}$ exhibit much more broadband response, as illustrated with numerically calculated spectra in Fig.~\ref{fig:CD_OR_all}(d). The proposed background cancellation still requires accurate balancing the two transmission channels using achiral reference measurements.

\section{Summary and discussion}

To summarize, we have developed an approach for generating local optical chirality near a periodic integrated waveguide through the interaction of two guided modes. In the symmetric waveguide, the required modal superposition must be prepared through the excitation conditions and remains chiral over a substantial propagation distance only near the modal degeneracy, resulting in an intrinsically narrow spectral range. The introduced waveguide asymmetry, by contrast, couples the two modes in quadrature, so that the states with opposite chirality become eigenmodes of the waveguide. This removes the need to impose the required phase relation through the excitation conditions and locks the sign of the layer-averaged chirality to the modal branch. The chiral eigenmodes therefore persist away from exact degeneracy, enabling operation over a substantially broader spectral range. In both cases, the chiral layer interacts with the nonzero local chirality in the accessible half-space, even when the chirality integrated over the entire waveguide cross-section vanishes. Thus, neither a globally chiral field distribution nor longitudinal twisting of the waveguide is required.

The perturbation theory connects these modal properties to experimentally accessible sensing observables. In the symmetric waveguide, differential transmission provides a signal linear in the chiral perturbation, providing a direct but narrow-band CD signature. The asymmetric waveguide supports chirality-carrying eigenmodes over a substantially broader spectral range, but modal dichroism produces a structural background in the differential transmission $\Delta T_a$. This background can be removed using a reference measurement, although the resulting signal generally contains contributions from both the real and imaginary parts of the chiral coupling. A different approach, based on comparing two waveguides with opposite rotations cancels the leading structural background and provides a more robust means of isolating the chiral response. The symmetric implementation therefore offers a comparatively direct but narrowband measurement, whereas the asymmetric implementation enables broadband sensing but requires compensation of the structural background.

\begin{figure}[t]
    \centering
    \includegraphics[width=\columnwidth]{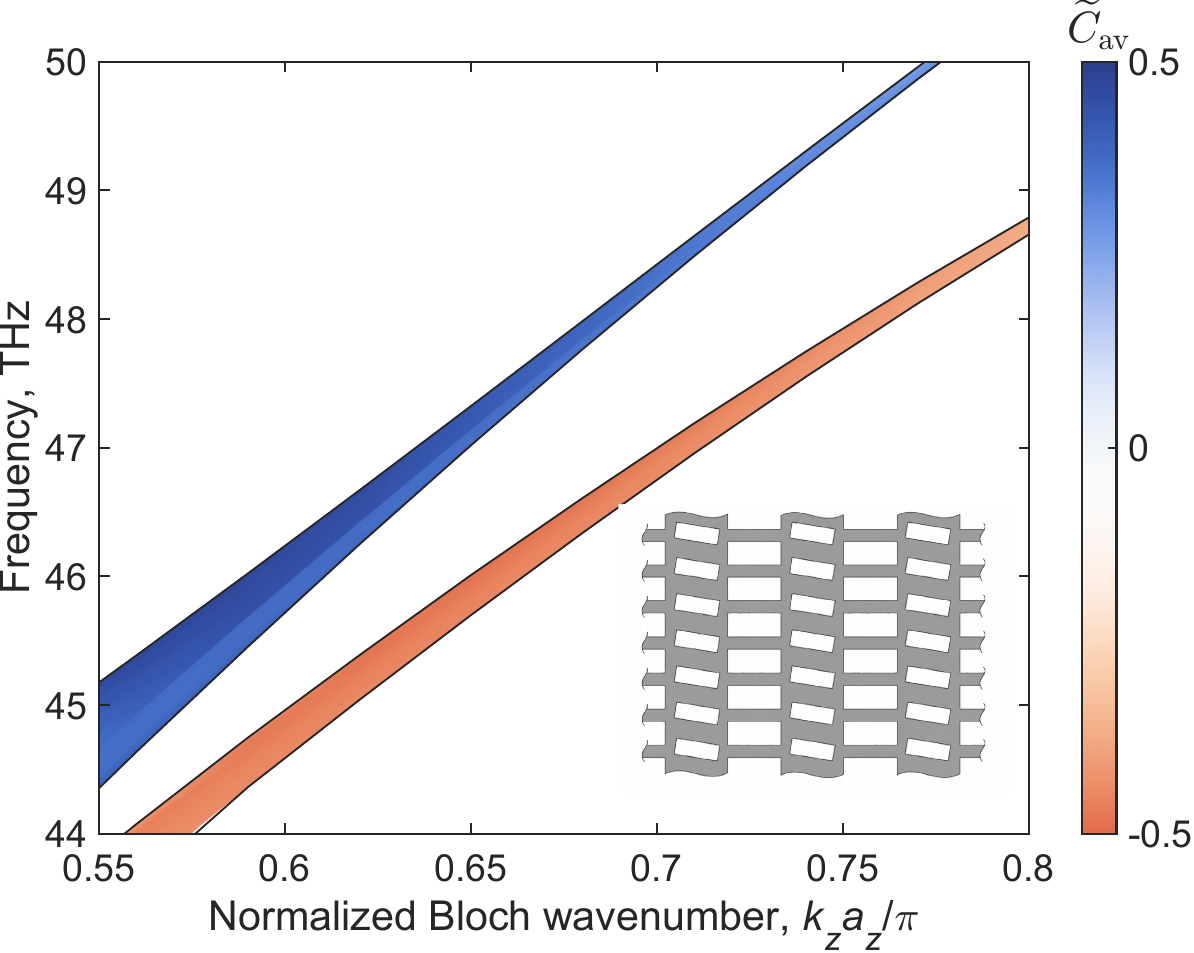}
    \caption{Projected band structure of the periodic waveguide array. For each
    of the two modes, the shaded region shows the frequency range obtained by
    varying the transverse Bloch wavenumber over
    $0\leq k_x d_x/\pi\leq 1$ at fixed longitudinal Bloch wavenumber
    $k_z d_z/\pi$. Period in transverse direction was taken $d_x=4.5$~$\mu$m. The color represents the normalized optical chirality averaged over a $500\,\mathrm{nm}$-thick layer above the array; red and blue
    denote negative and positive chirality, respectively.}
    \label{fig:array}
\end{figure}

A natural extension of this approach is an array of coupled two-mode waveguides, which would increase the area and volume interacting with the chiral layer. An array of symmetric waveguides, however, does not support ``chiral'' eigenmodes: the two relevant modal components enter the array Bloch modes with a relative phase of $0$ or $\pi$, resulting in zero chirality. This limitation can be avoided by constructing the array from identically oriented asymmetric waveguides, as schematically shown on the inset in Fig.~\ref{fig:array}. Provided that the coupling between neighboring waveguides remains weaker than the asymmetry-induced splitting, the opposite-chirality branches remain mostly decoupled. The resulting two-dimensional Bloch bands therefore retain nonzero layer-averaged chirality.
This behavior is visualized in Fig.~\ref{fig:array} by projecting the two-dimensional dispersion $\omega(k_x,k_z)$ onto the $(k_z,\omega)$ plane, such that the variation with $k_x$ produces finite-width bands whose color represents the average chirality. The two bands remain spectrally separated and retain opposite signs of the layer-averaged chirality throughout their entire widths, i.e., for all considered values of $k_x$. Thus, the inter-waveguide coupling produces the transverse dispersion of each band while leaving the opposite-chirality branches decoupled to leading order. Overall, the proposed mechanism provides a basis for developing spatially multiplexed waveguide sensors, as well as for interferometric or resonant structures designed to enhance the interaction of guided light with thin layers of chiral materials.

\section{Acknowledgments}
\begin{acknowledgments}
   The work was supported by the Russian Science Foundation, project \#25-72-10103, https://rscf.ru/project/25-72-10103/
   Authors thank Ivan Sinev and Denis Baranov for fruitful discussions.
\end{acknowledgments}

\bibliography{References.bib}

\newpage
\clearpage
\null
\thispagestyle{empty}

\onecolumngrid

\begin{center}
{\large\bfseries
Broadband chiroptical sensing with two-mode subwavelength grating waveguides:\\[4pt]
Supporting Information
}
\end{center}

\twocolumngrid

\section{I. Fields approximation}

The exact expression for the fields of the TE$_{01}$ and HE$_{11}^y$ of a cylindrical dielectric waveguide are given e.g. in the book~\cite{Okamoto}, Sec. 3.2.1--3.2.3. To simplify the exact field expressions we expand Macdonald functions for $\gamma r \gg 1$: $K_\alpha(z) \approx \sqrt{\dfrac{\pi}{2z}} e^{-z}$, and take $s\approx -1$.
\section{II. Perturbation theory for the periodic waveguide}


We consider a lossless periodic waveguide with period $d$ along the $z$ direction. At a fixed angular frequency $\omega$, the waveguide supports two nearly degenerate modes satisfying $(\beta_2-\beta_1)d\ll 1$, where $\beta_1$ and $\beta_2$ are their propagation constants. The fields of the unperturbed modes are written as
\begin{equation}
\begin{aligned}
\VB{E}_n^{(0)} &= \VB{e}_n^{(0)}(\VB{r}) e^{i\beta_n z}, \\
\VB{H}_n^{(0)} &= \VB{h}_n^{(0)}(\VB{r}) e^{i\beta_n z},
\end{aligned}
\end{equation}
where $\VB{e}_n^{(0)}$ and $\VB{h}_n^{(0)}$ are periodic along $z$ with period $d$. Assuming that the small perturbations $\Delta\eps(\VB{r})$, $\Delta\mu(\VB{r})$, and $\chi(\VB{r})$ are also periodic along $z$, we seek the fields of the perturbed waveguide in the form
\begin{equation}
\begin{aligned}
\VB{E} &= \VB{e}\,e^{i\beta z}
= \left( a_1 \VB{e}_1^{(0)} + a_2 \VB{e}_2^{(0)} \right)e^{i\beta z}, \\
\VB{H} &= \VB{h}\,e^{i\beta z}
= \left( a_1 \VB{h}_1^{(0)} + a_2 \VB{h}_2^{(0)} \right)e^{i\beta z}.
\end{aligned}
\label{eq:ansatz}
\end{equation}

The constitutive relations for the unperturbed waveguide are
\begin{equation}
\begin{aligned}
\VB{D}_n^{(0)}(\VB{r})
&= \eps_0\eps^{(0)}(\VB{r})\VB{E}_n^{(0)}(\VB{r}), \\
\VB{B}_n^{(0)}(\VB{r})
&= \mu_0\mu^{(0)}(\VB{r})\VB{H}_n^{(0)}(\VB{r}),
\end{aligned}
\end{equation}
whereas those for the perturbed waveguide containing the chiral layer are
\begin{equation}
\begin{aligned}
\VB{D}(\VB{r})
&= \eps_0\left[\eps^{(0)}(\VB{r})+\Delta\eps(\VB{r})\right]\VB{E}(\VB{r})
+ i\frac{\chi(\VB{r})}{c}\VB{H}(\VB{r}), \\
\VB{B}(\VB{r})
&= \mu_0\left[\mu^{(0)}(\VB{r})+\Delta\mu(\VB{r})\right]\VB{H}(\VB{r})
- i\frac{\chi(\VB{r})}{c}\VB{E}(\VB{r}).
\end{aligned}
\end{equation}

Consider the vector identities
\begin{equation}
\begin{aligned}
\nabla \cdot \left(\VB{E} \times \VB{H}_m^{(0)*}\right)
&= \VB{H}_m^{(0)*} \cdot \left(\nabla \times \VB{E}\right)
- \VB{E} \cdot \left(\nabla \times \VB{H}_m^{(0)*}\right), \\
\nabla \cdot \left(\VB{E}_m^{(0)*} \times \VB{H}\right)
&= \VB{H} \cdot \left(\nabla \times \VB{E}_m^{(0)*}\right)
- \VB{E}_m^{(0)*} \cdot \left(\nabla \times \VB{H}\right).
\end{aligned}
\label{eq:Pert_relations}
\end{equation}
Summing these identities and simplifying their right-hand sides using Maxwell's equations yields
\begin{equation}
\begin{aligned}
& \nabla \cdot \left( (\VB{e} \times \VB{h}_m^{(0)*} + \VB{e}_m^{(0)*} \times \VB{h})e^{i(\beta-\beta_m)z} \right) = \\
= & i\omega \left( \eps_0\Delta\eps\VB{e}_m^{(0)*} \cdot  \VB{e}
+ \mu_0\Delta\mu\VB{h}_m^{(0)*} \cdot  \VB{h} + \right. \\
+ & \left. i\dfrac{\chi}{c}(\VB{e}_m^{(0)*} \cdot \VB{h} - \VB{e} \cdot \VB{h}_m^{(0)*})
\right)e^{i(\beta-\beta_m)z}.
\end{aligned}
\label{eq:reciprocity}
\end{equation}

We substitute the ansatz for the perturbed fields, Eq.~\eqref{eq:ansatz}, and integrate Eq.~\eqref{eq:reciprocity} over one unit cell (UC), which is chosen without loss of generality as $0\leq z\leq d$. Applying the divergence theorem to the left-hand side and using the periodicity of the Bloch envelopes gives
\begin{equation}
LHS =
\left(e^{i(\beta-\beta_m)d}-1\right)
\sum_n a_n S_{mn},
\label{eq:reciprocity_integrated}
\end{equation}
where
\begin{equation}
S_{mn}
=
\int_S
\left(
\VB{e}_n^{(0)} \times \VB{h}_m^{(0)*}
+
\VB{e}_m^{(0)*} \times \VB{h}_n^{(0)}
\right)\cdot\hat{\VB{z}}\,dS.
\label{eq:Fm(z)}
\end{equation}
Here, the contributions from the transverse boundaries vanish because the guided-mode fields decay in the transverse directions.

For a weak perturbation, $|(\beta-\beta_m)d|\ll 1$, and hence
\begin{equation}
e^{i(\beta-\beta_m)d}-1
\approx i(\beta-\beta_m)d.
\end{equation}
Therefore,
\begin{equation}
LHS =
i(\beta-\beta_m)d
\sum_n a_n S_{mn}.
\label{eq:main_balance}
\end{equation}

Due to opposite parity with respect to transverse $x=0$ plnae, the unperturbed modes satisfy
\begin{equation}
S_{mn}=0,
\qquad m\neq n,
\end{equation}
while the diagonal terms are
\begin{equation}
S_{nn}=4P_n,
\label{eq:Snn}
\end{equation}
where $P_n$ is the power carried by the $n$th mode through a transverse plane. Thus,
\begin{equation}
LHS =
i(\beta-\beta_m)d\,a_mS_{mm}.
\end{equation}

After substituting Eq.~\eqref{eq:ansatz}, the right-hand side of Eq.~\eqref{eq:reciprocity} takes the form
\begin{equation}
\begin{aligned}
RHS
=
i\omega\sum_n a_n
\int_{UC}
\left[
\eps_0\Delta\eps\,
\VB{e}_m^{(0)*}\cdot\VB{e}_n^{(0)}
+
\mu_0\Delta\mu\,
\VB{h}_m^{(0)*}\cdot\VB{h}_n^{(0)}
\right.
\\
\left.
+
i\frac{\chi}{c}
\left(
\VB{e}_m^{(0)*}\cdot\VB{h}_n^{(0)}
-
\VB{e}_n^{(0)}\cdot\VB{h}_m^{(0)*}
\right)
\right]
e^{i(\beta-\beta_m)z}\,dV.
\end{aligned}
\end{equation}
Within one unit cell, the phase factor can be approximated as $e^{i(\beta-\beta_m)z}\approx 1$. Equating the left- and right-hand sides, dividing them by $ik_0dS_{mm}$, and taking into account Eq.~\eqref{eq:Snn} gives
\begin{equation}
\sum_n
\left[
(\tilde{\beta}-\tilde{\beta}_m)\delta_{mn}-V_{mn}
\right]a_n=0,
\end{equation}
where $\tilde{\beta} = \beta/k_0$, $\tilde{\beta}_m = \beta_m/k_0$,
\begin{equation}
\begin{aligned}
{V}_{mn}
=
\frac{1}{4 P_m d}
\int_{UC}
\left[
\dfrac{1}{Z_0}\Delta\eps\,
\VB{e}_m^{(0)*}\cdot\VB{e}_n^{(0)}
+
Z_0\Delta\mu\,
\VB{h}_m^{(0)*}\cdot\VB{h}_n^{(0)}
\right.
\\
\left.
+
i\chi
\left(
\VB{e}_m^{(0)*}\cdot\VB{h}_n^{(0)}
-
\VB{e}_n^{(0)}\cdot\VB{h}_m^{(0)*}
\right)
\right]\,dV,
\end{aligned}
\label{eq:Vmn}
\end{equation}
$Z_0=\sqrt{\mu_0/\eps_0}$ is the free space impedance. 
Consequently, the propagation constants of the perturbed modes are determined by
\begin{equation}
\det
\begin{pmatrix}
\tilde\beta-\tilde\beta_1-V_{11} & -V_{12} \\
-V_{21} & \tilde\beta-\tilde\beta_2-V_{22}
\end{pmatrix}
=0.
\label{eq:SI_disp_eq}
\end{equation}
In the main text we use constant values of $V_{mn}$ calculated at the degeneracy point, and dispersions $\beta_{1,2}(\omega)$ are linearized near the degeneracy point.




The chirality-induced mode splitting predicted by the perturbation theory described above was verified using full-wave simulations in COMSOL Multiphysics. For the perturbative calculations, we first numerically computed the eigenmodes of the lossless unperturbed waveguide. Particular care was taken to preserve the mirror symmetry of the numerical model when constructing the mesh~\cite{LeeSciRep2018}. Specifically, the mesh was made symmetric with respect to the vertical mirror plane to prevent spurious mode splitting caused by discretization-induced symmetry breaking. The two unperturbed modes were chosen to have opposite parity under reflection in this plane. The coupling constants were then evaluated using Eq.~\eqref{eq:Vmn} and used to calculate the mode splitting with Eq.~\eqref{eq:SI_disp_eq}.

For the full-wave calculations, the chiral layer with real-valued $\chi$ was explicitly included in the waveguide model, and the electromagnetic wave equations incorporating its constitutive relations were implemented directly in COMSOL. The resulting eigenfrequency splitting, $\Delta\omega=|\omega_1-\omega_2|$, was converted into an approximate propagation-constant splitting using
\begin{equation}
\Delta\beta
\approx \dfrac{\Delta\omega}{\sqrt{
v_{g_1}v_{g_2}}},
\end{equation}
where $v_{g_1}$ and $v_{g_2}$ are the group velocities of the unperturbed modes at the degeneracy point.

\begin{figure}[t]
    \centering
    \includegraphics[width=0.99\linewidth]{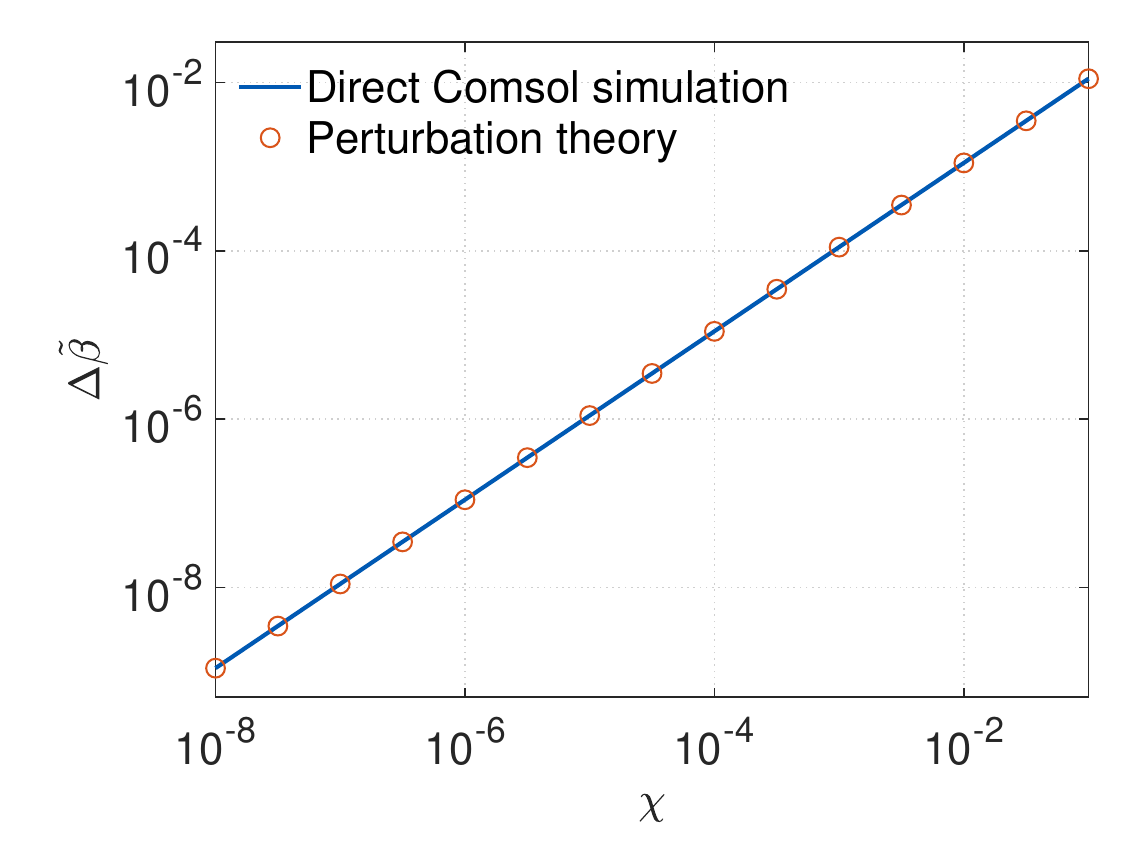}
    \caption{Normalized propagation-constant splitting $\Delta\tilde{\beta}$ as a function of the real-valued chirality parameter $\chi$ for a periodic waveguide coated with a 500-nm-thick chiral layer. The parameters of the waveguide are given in the main text. The blue solid line represents full-wave COMSOL simulations; the orange circles correspond to first-order
    perturbation theory.}
    \label{fig:chiral_splitting}
\end{figure}

Figure~\ref{fig:chiral_splitting} compares the normalized propagation constant splitting, $\Delta\tilde{\beta}$, obtained using the two approaches as a function of $\chi$. The full-wave results (blue line) closely agree with the perturbative predictions (orange circles) throughout the range considered. The splitting scales linearly with $\chi$ over several orders of magnitude, consistent with the first-order perturbative description of the chiral interaction.

\section{III. Symmetry properties of the perturbations}
\label{sec:perturbation_symmetry}

We use the modes of the lossless symmetric waveguide as the unperturbed basis and consider three perturbations: material absorption, the Pasteur coupling of the chiral layer, and the geometrical asymmetry produced by rotation of the holes about the $y$ axis. To determine the structure of the corresponding coupling matrices, we first summarize the relevant symmetry properties of the unperturbed modes.

Let $R_x\operatorname{diag}(-1,1,1)$ and $R_z\operatorname{diag}(1,1,-1)$ denote the matrices representing reflections with respect to the $x=0$ and $z=0$ planes, respectively. Since the electric field is a polar vector and the magnetic field is an axial vector, their transformations under mirror operator $\hat{\mathcal{M}}_x$ are
\begin{equation}
\begin{aligned}
\left(\hat{\mathcal{M}}_x\VB{e}\right)(\VB{r})
&=R_x\VB{e}(R_x\VB{r}),\\
\left(\hat{\mathcal{M}}_x\VB{h}\right)(\VB{r})
&=-R_x\VB{h}(R_x\VB{r}).
\end{aligned}
\label{eq:Mx_field_operator}
\end{equation}

Because the lossless symmetric waveguide is invariant under $\hat{\mathcal{M}}_x$, its nondegenerate modes can be chosen as eigenstates of this operator with eigenvalues of the $n$th mode denoted by $p_n=\pm1$. The two modes considered here have opposite mirror parities, so that $p_1p_2=-1$. Using Eq.~\eqref{eq:Mx_field_operator} and $R_x^{-1}=R_x$, the eigenvalue equation can be written explicitly as
\begin{equation}
\begin{aligned}
\VB{e}_n^{(0)}(R_x\VB{r})
&=p_n R_x\VB{e}_n^{(0)}(\VB{r}),\\
\VB{h}_n^{(0)}(R_x\VB{r})
&=-p_n R_x\VB{h}_n^{(0)}(\VB{r}).
\end{aligned}
\label{eq:Mx_mode_symmetry}
\end{equation}

Reflection $\hat{\mathcal{M}}_z$ reverses the propagation direction and therefore transforms a mode with propagation constant $\beta_n$ into a mode with $-\beta_n$. Time reversal $\mathcal{T}$ also reverses the propagation direction and, for the $e^{-i\omega t}$ convention, acts as $\VB{e}\rightarrow\VB{e}^{*}$ and $\VB{h}\rightarrow-\VB{h}^{*}$. Consequently, the combined antiunitary operator $\hat{\mathcal{M}}_z\mathcal{T}$ leaves the propagation constant unchanged and acts on the fields as
\begin{equation}
\begin{aligned}
\left(\hat{\mathcal{M}}_z\mathcal{T}\VB{e}\right)(\VB{r})
&=R_z\VB{e}^{*}(R_z\VB{r}),\\
\left(\hat{\mathcal{M}}_z\mathcal{T}\VB{h}\right)(\VB{r})
&=R_z\VB{h}^{*}(R_z\VB{r}).
\end{aligned}
\label{eq:MzT_field_operator}
\end{equation}
The minus sign introduced by time reversal for the magnetic field is canceled by the axial-vector transformation under $\hat{\mathcal{M}}_z$. The global phase of each nondegenerate mode can be chosen such that its eigenvalue under $\hat{\mathcal{M}}_z\mathcal{T}$ is equal to unity. Equation~\eqref{eq:MzT_field_operator} then gives
\begin{equation}
\begin{aligned}
\VB{e}_n^{(0)}(R_z\VB{r})
&=R_z\VB{e}_n^{(0)*}(\VB{r}),\\
\VB{h}_n^{(0)}(R_z\VB{r})
&=R_z\VB{h}_n^{(0)*}(\VB{r}).
\end{aligned}
\label{eq:MzT_mode_symmetry}
\end{equation}
These relations apply to the lossless reciprocal reference waveguide.


Equations~\eqref{eq:Mx_mode_symmetry} and \eqref{eq:MzT_mode_symmetry} determine which matrix elements are allowed and fix their phases. For $m\neq n$, the scalar products $\VB{e}_m^{(0)*}\cdot\VB{e}_n^{(0)}$ and $\VB{h}_m^{(0)*}\cdot\VB{h}_n^{(0)}$ are odd under $\hat{\mathcal{M}}_x$. Consequently, a scalar perturbation that is even under $\hat{\mathcal{M}}_x$ does not couple the two modes through the electric--electric or magnetic--magnetic overlap integrals. By contrast,
\begin{equation}
F_{12}(\VB{r})=
\VB{e}_1^{(0)*}\cdot\VB{h}_2^{(0)}
-\VB{e}_2^{(0)}\cdot\VB{h}_1^{(0)*}
\label{eq:chiral_cross_integrand}
\end{equation}
is even under $\hat{\mathcal{M}}_x$, whereas the corresponding diagonal quantities $F_{nn}$ are odd. In addition, Eq.~\eqref{eq:MzT_mode_symmetry} gives
\begin{equation}
F_{12}(R_z\VB{r})=F_{12}^{*}(\VB{r}).
\end{equation}
Thus, for an integration region symmetric under $\hat{\mathcal{M}}_z$, the integral of $F_{12}$ is real in the phase convention of Eq.~\eqref{eq:MzT_mode_symmetry}.

The chirality of individual modes, 
\begin{equation}
C_{n}(\VB{r}) \propto
\VB{e}_n^{(0)*}\cdot\VB{h}_n^{(0)}
-\VB{e}_n^{(0)}\cdot\VB{h}_n^{(0)*},
\end{equation}
vanishes for an integration region symmetric under $\hat{\mathcal{M}}_x$.

We first consider material absorption represented by imaginary-valued perturbations $\Delta\eps=i\Delta\eps''$ and $\Delta\mu=i\Delta\mu''$. We assume that their spatial distributions are even under $\hat{\mathcal{M}}_x$. Their cross-coupling matrix elements then vanish, while their diagonal overlaps are real before multiplication by $i$. Therefore, absorption produces purely imaginary diagonal corrections,
\begin{equation}
V_{nn}^{(loss)}=i\gamma_n,
\; V_{12}^{(loss)}=V_{21}^{(loss)}=0.
\label{eq:loss_matrix_elements}
\end{equation}
The real-valued quantities $\gamma_n\in\mathbb{R}$ describe the generally different attenuation of the individual modes.

For the Pasteur perturbation, the diagonal matrix elements vanish because $F_{nn}$ is odd under $\hat{\mathcal{M}}_x$. The cross overlap is real, and the explicit factor $i$ in the Pasteur contribution gives
\begin{equation}
V_{12}^{(ch)}=i\delta,
\;
V_{21}^{(ch)}=-i\delta,
\label{eq:chiral_matrix_elements}
\end{equation}
where $\delta=\chi\kappa$ and $\kappa\in\mathbb{R}$ contains the real-valued field overlap and normalization factors. Hence, the chiral perturbation produces antisymmetric nondiagonal matrix elements. Although $\kappa$ is real, $\delta$ is generally complex because the Pasteur parameter $\chi$ can be complex.

Finally, let $\Delta\eps_{rot}$ denote the first-order geometrical perturbation produced by rotation of the holes around $y$ axis. Reflection with respect to either the $x=0$ or $z=0$ plane changes the direction of rotation, so that
\begin{equation}
\Delta\eps_{rot}(R_x\VB{r})
=-\Delta\eps_{rot}(\VB{r}),
\;
\Delta\eps_{rot}(R_z\VB{r})
=-\Delta\eps_{rot}(\VB{r}).
\label{eq:rotation_perturbation_symmetry}
\end{equation}
The diagonal electric overlaps are even under $\hat{\mathcal{M}}_x$ and therefore give zero when multiplied by $\Delta\eps_{rot}$. The cross electric overlap is odd under $\hat{\mathcal{M}}_x$, so its product with $\Delta\eps_{rot}$ is even and the modes can couple. At the same time, Eq.~\eqref{eq:MzT_mode_symmetry} yields
\begin{equation}
\left[
\VB{e}_1^{(0)*}\cdot\VB{e}_2^{(0)}
\right](R_z\VB{r})
=
\left[
\VB{e}_1^{(0)*}\cdot\VB{e}_2^{(0)}
\right]^*(\VB{r}).
\end{equation}
Together with the second relation in Eq.~\eqref{eq:rotation_perturbation_symmetry}, this shows that the rotation-induced cross overlap is purely imaginary. Consequently,
\begin{equation}
V_{12}^{(rot)}=i\delta_{rot},
\;
V_{21}^{(rot)}=-i\delta_{rot},
\label{eq:rotation_matrix_elements}
\end{equation}
and real-valued $\delta_{rot}\in\mathbb{R}$ changes sign when the direction of rotation is reversed.

To first order, the three perturbations are additive and the coupled-mode equation becomes
\begin{equation}
\det
\begin{pmatrix}
\tilde{\beta}_c-\tilde{\beta}_1-i\gamma_1
&-i\left(\delta_{rot}+\delta\right)\\
i\left(\delta_{rot}+\delta\right)
&\tilde{\beta}_c-\tilde{\beta}_2-i\gamma_2
\end{pmatrix}
=0.
\label{eq:CMT_all_perturbations}
\end{equation}
Thus, mirror-symmetric absorption produces imaginary diagonal corrections, whereas chirality and geometrical asymmetry produce antisymmetric nondiagonal corrections. The geometrical coefficient $\delta_{rot}$ is real, while the chiral coefficient $\delta=\chi\kappa$ inherits the generally complex phase of $\chi$.

At second order in the rotation angle, the effective rotation-induced coupling remains of the form $V_{12}^{(rot)}=i\delta_{rot}$ with real $\delta_{rot}$. Even-order terms in the geometrical perturbation are even under $\hat{\mathcal{M}}_x$ and therefore produce only real diagonal corrections for the two modes. These small second-order diagonal shifts are neglected in the present coupled-mode model. Mixed loss--rotation, loss--chirality, and rotation--chirality corrections are also beyond the adopted first-order approximation.
\section{IV. Transmission characteristics of a symmetric waveguide}

The coupled-mode equation describing the modes of a symmetric waveguide covered with a chiral layer is:
\begin{equation}
\det
\begin{pmatrix}
\tilde{\beta}_c-\tilde{\beta}^{(corr)}_1 & -i\delta \\
i\delta & \tilde{\beta}_c-\tilde{\beta}^{(corr)}_2
\end{pmatrix}
=0,
\label{eq:M_simplified}
\end{equation}
where $\delta \propto \chi$ describes the chiral perturbation and $\tilde{\beta}^{(corr)}_n = \tilde{\beta}_n' + i\tilde{\beta}_n'' + i\gamma_n^{ch}$ includes losses of the waveguide $\tilde{\beta}_n''$ as well as achiral losses due to chiral layer $\gamma_n^{ch}$. The propagation constants of the coupled modes are
\begin{equation}
\tilde{\beta}_c=\beta_s\pm\alpha,
\label{eq:beta_coupled}
\end{equation}
where
\begin{equation}
\begin{aligned}
&\beta_s=\frac{\tilde{\beta}^{(corr)}_1+\tilde{\beta}^{(corr)}_2}{2},
\\
&\alpha=\sqrt{\frac{\Delta\beta^2}{4}+\delta^2},
\\
&\Delta\beta=\tilde{\beta}^{(corr)}_1-\tilde{\beta}^{(corr)}_2.
\end{aligned}
\end{equation}
The corresponding eigenvectors are
\begin{equation}
v_1=
\begin{pmatrix}
i\\
\eps
\end{pmatrix},
\qquad
v_2=
\begin{pmatrix}
\eps\\
i
\end{pmatrix},
\label{eq:vectors_coupled}
\end{equation}
where
\begin{equation}
\eps=\frac{\alpha-\Delta\beta/2}{\delta}
=\frac{\delta}{\alpha+\Delta\beta/2}.
\end{equation}

Consider a signal $(c_1,c_2)^T$, where $c_1$ and $c_2$ are the amplitudes of the two modes, launched into a waveguide covered with a chiral layer. The normalized length of the layer is $L=k_0l$, where $l$ is its physical length. The amplitudes at the output are
\begin{equation}
\begin{pmatrix}
c_{1,out}\\
c_{2,out}
\end{pmatrix}
=SPS^{-1}
\begin{pmatrix}
c_1\\
c_2
\end{pmatrix},
\end{equation}
where
\begin{equation}
S=
\begin{pmatrix}
i & \eps\\
\eps & i
\end{pmatrix},
\qquad
S^{-1}=-\frac{1}{1+\eps^2}
\begin{pmatrix}
i & -\eps\\
-\eps & i
\end{pmatrix},
\end{equation}
and
\begin{equation}
P=
\begin{pmatrix}
e^{i(\beta_s+\alpha)L} & 0\\
0 & e^{i(\beta_s-\alpha)L}
\end{pmatrix}.
\end{equation}

The input and output powers are $P_{in}=|c_1|^2+|c_2|^2$ and $P_{out}=|c_{1,out}|^2+|c_{2,out}|^2$, respectively. For the $(1,0)^T$ and $(0,1)^T$ inputs, the output amplitudes are
\begin{equation}
\begin{aligned}
\begin{pmatrix}
c_{1,out}\\
c_{2,out}
\end{pmatrix}_{(1,0)}
&=\frac{1}{1+\eps^2}
\begin{pmatrix}
e^{i(\beta_s+\alpha)L}+\eps^2e^{i(\beta_s-\alpha)L}\\
i\eps\left[e^{i(\beta_s-\alpha)L}-e^{i(\beta_s+\alpha)L}\right]
\end{pmatrix},\\[2mm]
\begin{pmatrix}
c_{1,out}\\
c_{2,out}
\end{pmatrix}_{(0,1)}
&=\frac{1}{1+\eps^2}
\begin{pmatrix}
i\eps\left[e^{i(\beta_s+\alpha)L}-e^{i(\beta_s-\alpha)L}\right]\\
\eps^2e^{i(\beta_s+\alpha)L}+e^{i(\beta_s-\alpha)L}
\end{pmatrix}.
\end{aligned}
\end{equation}

The modal power conversion is therefore
\begin{equation}
R_s=e^{-2\beta_s''L}\frac{|\eps|^2}{|1+\eps^2|^2}
\left[e^{-2\alpha''L}+e^{2\alpha''L}-2\cos(2\alpha'L)\right].
\label{eq:Rs_exact}
\end{equation}

For the $(1,\pm i)^T/\sqrt{2}$ inputs, the output amplitudes are
\begin{equation}
\begin{aligned}
&\begin{pmatrix}
c_{1,out}\\
c_{2,out}
\end{pmatrix}_{\pm}
=\frac{1}{\sqrt{2}(1+\eps^2)}\cdot \\
&\cdot \begin{pmatrix}
(1\mp\eps)e^{i(\beta_s+\alpha)L}
+\eps(\pm1+\eps)e^{i(\beta_s-\alpha)L}\\
-i\eps(1\mp\eps)e^{i(\beta_s+\alpha)L}
+i(\pm1+\eps)e^{i(\beta_s-\alpha)L}
\end{pmatrix}.
\end{aligned}
\end{equation}
The differential transmission $\Delta T_s=T_+-T_-$ is then
\begin{equation}
\begin{aligned}
\Delta T_s=\frac{2e^{-2\beta_s''L}}{|1+\eps^2|^2}
& \left[
\eps'(1+|\eps|^2) \left(e^{2\alpha''L}-e^{-2\alpha''L}\right) + \right.\\
& \left. + 2\eps''(1-|\eps|^2)\sin(2\alpha'L)
\right].
\label{eq:DeltaTs_exact}
\end{aligned}
\end{equation}

To simplify the obtained expressions we use weak accumulated chiral coupling approximation, $|\delta|L \ll 1$, which well agrees with realistic conditions: 
\begin{equation}
\begin{aligned}
R_s &\approx
\dfrac{4e^{-2\beta_s''L}|\delta|^2}{|\Delta\beta|^2}
\left[
\sin^2\left(\dfrac{\Delta\beta'L}{2}\right)
+
\sinh^2\left(\dfrac{\Delta\beta''L}{2}\right)
\right],\\
\Delta T_s &\approx
\dfrac{4e^{-2\beta_s''L}}{|\Delta\beta|^2}
\left[
(\delta'\Delta\beta'+\delta''\Delta\beta'')
\sinh(\Delta\beta''L) + \right.\\
& \left. +
(\delta''\Delta\beta'-\delta'\Delta\beta'')
\sin(\Delta\beta'L)
\right].
\end{aligned}
\label{eq:Rs_DeltaTs_weak_delta}
\end{equation}
Expanding these expressions assuming $|\Delta\beta|L \ll 1$ yields:
\begin{equation}
\begin{aligned}
R_s &\approx
e^{-2\beta_s''L}|\delta|^2L^2
\left[
1+
\dfrac{\left(\Delta\beta''^2-\Delta\beta'^2\right)L^2}{12}
\right],\\
\Delta T_s &\approx
4e^{-2\beta_s''L}L
\left[
\delta'' + \right.\\
& \left. + \dfrac{L^2}{6}
\left(
\delta'\Delta\beta'\Delta\beta''
+
\delta''
\left[\Delta\beta''^2-\Delta\beta'^2\right]
\right)
\right],
\end{aligned}
\label{eq:Rs_DeltaTs_small_detuning_SI}
\end{equation}

One can observe that $R_s$ always exhibits maximum at the degeneracy point $\Delta\beta'=0$ regardless of modal dichroism $\Delta\beta''$. Away from the degeneracy point $R_s$ decreases $\propto \Delta\beta'^2$, reducing significantly for spectral detuning $\Delta\beta' \gtrsim 1/L$.

On the other hand, $\Delta T_s$ contains term proportional to $\Delta\beta'\Delta\beta''\delta'$, producing an asymmetric, locally linear spectral contribution and mixing the real and imaginary parts of the chiral coupling. This term shift the maximum of $\Delta T_s$ from the degeneracy point by the value $\propto \left|\dfrac{\delta'}{\delta''}\Delta\beta''\right|$. This shift is negligible as soon as $|\Delta\beta''| L \ll \left|\dfrac{\delta''}{\delta'}\right|$. In this case $\Delta T_s$ also exhibits parabolic spectral response similar to the $R_s$.

\section{V. Transmission characteristics of an asymmetric waveguide}

We first consider an asymmetric waveguide with $\chi=0$, which is described by the coupled-mode matrix
\begin{equation}
M_0=
\begin{pmatrix}
\beta_1 & i\delta_{rot}\\
-i\delta_{rot} & \beta_2
\end{pmatrix},
\end{equation}
where $\beta_1$ and $\beta_2$ are generally complex. For convenience we introduce
\begin{equation*}
\begin{aligned}
\beta_s=\frac{\beta_1+\beta_2}{2},
\;
\Delta\beta=\beta_1-\beta_2,
\\
\alpha_0=\sqrt{\frac{\Delta\beta^2}{4}+\delta_{rot}^2},\;
\eps_0=\frac{\alpha_0-\Delta\beta/2}{\delta_{rot}}
\end{aligned}
\end{equation*}
The propagation constants and the corresponding normalized eigenvectors are then
\begin{equation*}
\beta_{a_{1,2}}^{(0)}=\beta_s\pm\alpha_0,
\end{equation*}
\begin{equation*}
a_1=\frac{1}{\sqrt{1+|\eps_0|^2}}
\begin{pmatrix}
i\\
\eps_0
\end{pmatrix},
\;
a_2=\frac{1}{\sqrt{1+|\eps_0|^2}}
\begin{pmatrix}
\eps_0\\
i
\end{pmatrix}.
\end{equation*}
Because $\Delta\beta$ and $\eps_0$ are generally complex, the two right eigenvectors are not orthogonal under the conventional Hermitian inner product.

The chiral layer changes the off-diagonal coupling according to
\begin{equation}
\delta_{as}=\delta_{rot}+\delta,
\end{equation}
where $\delta$ is the coupling induced by the chiral layer. We assume that the diagonal propagation constants, including their imaginary parts, are unchanged in the covered section. Additional weak losses that are similar for both modes can be included in $\beta_s''$.

The propagation-constant splitting and the eigenvector parameter in the covered section are
\begin{equation}
\alpha=\sqrt{\frac{\Delta\beta^2}{4}+\delta_{as}^2},
\qquad
\eps=\frac{\alpha-\Delta\beta/2}{\delta_{as}}.
\end{equation}
The propagation matrix through a covered section with normalized length $L$ is
\begin{equation}
U=
\frac{1}{1+\eps^2}
\begin{pmatrix}
e^+ +\eps^2e^-
& i\eps\left[e^+ - e^- \right]\\
i\eps\left[e^- -e^+ \right]
& \eps^2e^+ + e^-
\end{pmatrix},
\label{eq:lossy_asymmetric_propagation_matrix}
\end{equation}
where $e^{\pm} = e^{i(\beta_s  \pm \alpha)L}$.

When the waveguide mode $a_j$ of the input waveguide is launched, the output state is
\begin{equation*}
c_{out}^{(j)}=U a_j,
\end{equation*}
and its total transmission is
\begin{equation}
T_{a_j}=\left(c_{out}^{(j)}\right)^{\dagger}c_{out}^{(j)}.
\label{eq:lossy_asymmetric_transmission_general}
\end{equation}
In terms of the matrix elements of $U$, the two transmissions are
\begin{equation}
\begin{aligned}
T_{a_1}&=\frac{
\left|iU_{11}+\eps_0U_{12}\right|^2
+\left|iU_{21}+\eps_0U_{22}\right|^2
}{1+|\eps_0|^2},\\
T_{a_2}&=\frac{
\left|\eps_0U_{11}+iU_{12}\right|^2
+\left|\eps_0U_{21}+iU_{22}\right|^2
}{1+|\eps_0|^2}.
\end{aligned}
\label{eq:lossy_asymmetric_transmissions}
\end{equation}

Equations~\eqref{eq:lossy_asymmetric_propagation_matrix}--\eqref{eq:lossy_asymmetric_transmissions} give the exact result within the two-mode model and can be evaluated directly without assuming weak modal dichroism or weak chiral coupling.

For a weak chiral perturbation relative to the geometrical rotation coupling, $|\delta|\ll|\delta_{rot}|$, the propagation matrix can be expanded as
\begin{equation}
U(\delta_{rot}+\delta)
=U_0
+\delta U_0'
+O(|\delta|^2),
\label{eq:U_first_order_delta}
\end{equation}
where $U_0=U(\delta_{rot}),
\qquad
U_0'=\left.
\frac{\partial U}{\partial\delta_{as}}
\right|_{\delta_{as}=\delta_{rot}}$.

To first order in $\delta$, the transmission of mode $a_j$ is
\begin{equation}
\begin{aligned}
T_{a_j}={}&T_{a_j}^{(0)}
+2\operatorname{Re}\left\{
\delta
\left(U_0a_j\right)^{\dagger}
U_0'a_j
\right\}
+O(|\delta|^2),
\end{aligned}
\label{eq:Taj_first_order_delta}
\end{equation}
where $T_{a_1}^{(0)}=e^{-2(\beta_s''+\alpha_0'')L},
\;\;
T_{a_2}^{(0)}=e^{-2(\beta_s''-\alpha_0'')L}$.
Consequently, differential transmission
\begin{equation}
\begin{aligned}
&\Delta T_a \equiv T_{a_1}-T_{a_2} = \Delta T_a^{(0)} + \\
&+2\operatorname{Re}\left\{
\delta\left[
\left(U_0a_1\right)^{\dagger}
U_0'a_1
-
\left(U_0a_2\right)^{\dagger}
U_0'a_2
\right]
\right\}
+O(|\delta|^2).
\end{aligned}
\label{eq:DeltaTa_first_order_lossy}
\end{equation}
contains differential transmission background
\begin{equation}
\Delta T_a^{(0)} =  -2e^{-2\beta_s''L}\sinh(2\alpha_0''L),
\end{equation}
produced by the modal dichroism $\Delta\beta'' \ne 0$. The background signal is independent of the chiral parameter and in the vicinity of the mode degeneracy, $|\Delta\beta'|\ll
2|\delta_{\mathrm{rot}}|$, it is linear in spectral detuning. Consequently, away from the degeneracy it can become rather strong. Such a background can be removed in two ways. The first one is to determine the background in a separate measurement and then subtract it. However, in this case the small signal is determined as the difference of two large signals measured in different experiments, which makes it challenging.

The alternative way to eliminate the unwanted background is to consider two waveguides with opposite rotations, corresponding to $\pm \delta_{rot}$ coupling constants. For the same corresponding mode $a_j$ in the two waveguides, the rotation-differential transmission is
\begin{equation}
\Delta T_r^{(j)}=
T_{+\theta,a_j}-T_{-\theta,a_j}.
\end{equation}
The two structures have identical transmission when $\delta=0$. Using their mirror relation and expanding to first order in $\delta$ gives
\begin{equation}
\Delta T_r^{(j)}=
4\operatorname{Re}\left\{
\delta
\left(U_0a_j\right)^{\dagger}
U_0'a_j
\right\}
+O(|\delta|^3).
\label{eq:DeltaTr_first_order_lossy}
\end{equation}

The derived expressions can be simplified using the following approximations:
\begin{equation}
    \begin{gathered}
        |\delta|L\ll 1,
        \;\;
        |\Delta\beta''|\ll \bar{\alpha}_0,\\
       |\alpha_0''|L \ll 1,
        \;\;
        |\delta_{\mathrm{rot}}|L\gg 1,
    \end{gathered}
    \label{eq:asymmetric-weak-loss-approximation}
\end{equation}
where $\bar{\alpha}_0=\sqrt{\delta_{\mathrm{rot}}^2+\frac{(\Delta\beta')^2}{4}}$,  $\alpha_0''\approx\frac{\Delta\beta'\Delta\beta''}{4\bar{\alpha}_0}$.
Such approximation is reasonable for the considered parameters: (1) the accumulated chiral correction is small, (2) modal dichroism weakly perturbs the splitting of the asymmetric modes, (3) differential attenuation accumulated by these modes remains small, (4) the length-enhanced propagation contribution dominates bounded mode-mismatch terms.

The simplified differential transmissions are:
\begin{equation}
\begin{aligned}
    &\Delta T_a^{(0)}
    \approx -4e^{-2\beta_s''L}\alpha_0''L,\\
    &\Delta T_a
    \approx
    -4e^{-2\beta_s''L}\left[ \alpha_0''L +  \dfrac{\delta_{\mathrm{rot}}}{\bar{\alpha}_0}L\left(
        \delta''
        -\frac{\alpha_0''}{\bar{\alpha}_0}\delta'
    \right) \right],\\
    &\Delta T_r^{(1,2)}
    \approx \mp 4 e^{-2\beta_s''L}\dfrac{\delta_{\mathrm{rot}}}{\bar{\alpha}_0}L \left(\delta''(1 \mp2\alpha_0''L) - \dfrac{\alpha_0''} {\bar{\alpha}_0}\delta'\right).
    \label{eq:DeltaT-weak-loss}
\end{aligned}
\end{equation}
The terms proportional to $\delta'$ in $\Delta T_a$ and $\Delta T_r$
are suppressed by the weak modal dichroism as compared to $\delta_{\mathrm{rot}}$.
For the parameters considered in the main text,
$\left|(\alpha_0''/\bar{\alpha}_0)(\delta'/\delta'')\right|
\lesssim 10^{-3}$ over the considered spectral range and can therefore
be neglected.

If one further considers not very large spectral detuning $\Delta\beta'\lesssim \delta_{\mathrm{rot}}$, the differential transmission are approximated as follows:
\begin{equation}
\begin{aligned}
    & \Delta T_a^{(0)}
    \approx
    -e^{-2\beta_s''L}\dfrac{\Delta\beta'\Delta\beta''}{|\delta_{\mathrm{rot}}|}L \zeta,\\
& \Delta T_a
    \approx
    \Delta T_a^{(0)} - 4e^{-2\beta_s''L}\dfrac{\delta_{\mathrm{rot}}}{{|\delta_{\mathrm{rot}}|}}
        \delta''
        L \zeta,\\
& \Delta T_r^{(1,2)} \approx \mp 4 e^{-2\beta_s''L}\dfrac{\delta_{\mathrm{rot}}}{{|\delta_{\mathrm{rot}}|}}\delta''L\, \left( 1 \mp \dfrac{\Delta\beta'\Delta\beta''}{2|\delta_{\mathrm{rot}}|}L \right)\zeta
\end{aligned}
\end{equation}
where $\zeta = \left( 1 - \dfrac{\Delta\beta'^2}{8\delta_{\mathrm{rot}}^2} \right)$ describes the parabolic spectral detuning. The leading contribution to the rotation-differential transmission $\Delta T_r^{(1,2)}$ is linearly proportional to $\delta''$ and $L$, even in $\Delta\beta'$, and maximal at the mode degeneracy. Its even spectral envelope decreases over the characteristic range $|\Delta\beta'|\sim2|\delta_{\mathrm{rot}}|$. Modal dichroism introduces an additional odd contribution proportional to $\Delta\beta'\Delta\beta''L$, making the spectrum asymmetric. This contribution originates from the expansion of the mode-dependent attenuation factor $e^{\mp2\alpha_0''L}$ and becomes non-negligible for sufficiently large spectral detuning, modal dichroism, or propagation length. Since $\alpha_0''$ is determined by the properties of the input waveguide with zero $\chi$, its effect can be calibrated using the corresponding measurements.

\end{document}